\documentclass[%
 preprint,
3p,times,onecolumn]{elsarticle}
\pdfoutput=1
\usepackage{etoolbox}
\usepackage[utf8]{inputenc}
\usepackage[colorlinks]{hyperref}
\usepackage{graphicx}
\usepackage{natbib}
\usepackage{tabularx}
\usepackage{lipsum}
\usepackage{wrapfig}
\usepackage{amsmath}
\usepackage{float}
\usepackage{subfigure}
\usepackage{gensymb}
\usepackage{hyperref}
\usepackage{cleveref}
\usepackage{chngcntr}
\usepackage{multirow}
\usepackage{makecell}
\usepackage{array}
\usepackage{booktabs}
\usepackage[none]{hyphenat} 
\usepackage{enumitem}
\usepackage{ragged2e}
\usepackage{amssymb}
\usepackage[section]{placeins}
\usepackage{siunitx}
\usepackage{epigraph}
\usepackage{lineno}
\usepackage{braket}
\usepackage{xparse,nameref}
\usepackage[usestackEOL]{stackengine}
\usepackage{xcolor}
\usepackage{mathtools}
\usepackage{multirow}
\usepackage{makecell}
\usepackage{array}
\usepackage{booktabs}
\usepackage{textcase}
\usepackage[toc,page]{appendix}
\usepackage{pdfpages}
\usepackage{titlesec}

\titleformat{\subsection} {\normalfont\bfseries}{\thesubsection}{1em}{}
\titleformat{\paragraph} {\normalfont\itshape}{\paragraph}{1em}{}
\usepackage{dcolumn}   

\newcolumntype{M}[1]{>{\centering\arraybackslash}m{#1}}

\usepackage[tablename=TABLE, labelsep=period, figurename=Figure, skip=0pt, font=small]{caption}

\makeatletter
\def\ps@pprintTitle{%
 \let\@oddhead\@empty
 \let\@evenhead\@empty
 \def\@oddfoot{}%
 \let\@evenfoot\@oddfoot}
\makeatother

\begin{document}
       
   \title{\textbf{Final results of Borexino on CNO solar neutrinos}}

\author[Milano]{D.~Basilico}
\author[Milano]{G.~Bellini}
\author[PrincetonChemEng]{J.~Benziger}
\author[LNGS]{R.~Biondi\fnref{MPI}}
\author[Milano]{B.~Caccianiga}
\author[Princeton]{F.~Calaprice}
\author[Genova]{A.~Caminata}
\author[Lomonosov]{A.~Chepurnov}
\author[Milano]{D.~D'Angelo}
\author[Peters]{A.~Derbin}
\author[LNGS]{A.~Di Giacinto}
\author[LNGS]{V.~Di Marcello}
\author[Princeton]{X.F.~Ding\fnref{IHEP}}
\author[Princeton]{A.~Di Ludovico\fnref{LNGSG}} 
\author[Genova]{L.~Di Noto}
\author[Peters]{I.~Drachnev}
\author[APC]{D.~Franco}
\author[Princeton,GSSI]{C.~Galbiati}
\author[LNGS]{C.~Ghiano}
\author[Milano]{M.~Giammarchi}
\author[Princeton]{A.~Goretti\fnref{LNGSG}}
\author[Lomonosov,Dubna]{M.~Gromov}
\author[Mainz]{D.~Guffanti\fnref{Bicocca}}
\author[LNGS]{Aldo~Ianni}
\author[Princeton]{Andrea~Ianni}
\author[Krakow]{A.~Jany}
\author[Kiev]{V.~Kobychev}
\author[London,Atomki]{G.~Korga}
\author[Juelich,RWTH]{S.~Kumaran\fnref{CALI}}
\author[LNGS]{M.~Laubenstein}
\author[Kurchatov,Kurchatovb]{E.~Litvinovich}
\author[Milano]{P.~Lombardi}
\author[Peters]{I.~Lomskaya}
\author[Juelich,RWTH]{L.~Ludhova}
\author[Kurchatov,Kurchatovb]{I.~Machulin}
\author[Mainz]{J.~Martyn}
\author[Milano]{E.~Meroni}
\author[Milano]{L.~Miramonti}
\author[Krakow]{M.~Misiaszek}
\author[Peters]{V.~Muratova}
\author[Kurchatov]{R.~Nugmanov}
\author[Munchen]{L.~Oberauer}
\author[Mainz]{V.~Orekhov}
\author[Perugia]{F.~Ortica}
\author[Genova]{M.~Pallavicini}
\author[Juelich,RWTH]{L.~Pelicci}
\author[Juelich]{\"O.~Penek\fnref{GSI}}
\author[Princeton]{L.~Pietrofaccia\fnref{LNGSG}}
\author[Peters]{N.~Pilipenko}
\author[UMass]{A.~Pocar}
\author[Kurchatov]{G.~Raikov}
\author[LNGS]{M.T.~Ranalli}
\author[Milano]{G.~Ranucci}
\author[LNGS]{A.~Razeto}
\author[Milano]{A.~Re}
\author[LNGS]{N.~Rossi}
\author[Munchen]{S.~Sch\"onert}
\author[Peters]{D.~Semenov}
\author[Juelich]{G.~Settanta\fnref{ISPRA}}
\author[Kurchatov,Kurchatovb]{M.~Skorokhvatov}
\author[Juelich,RWTH]{A.~Singhal}
\author[Dubna]{O.~Smirnov}
\author[Dubna]{A.~Sotnikov}
\author[LNGS]{R.~Tartaglia}
\author[Genova]{G.~Testera}
\author[Peters]{E.~Unzhakov}
\author[LNGS,Aquila]{F.L.~Villante}
\author[Dubna]{A.~Vishneva}
\author[Virginia]{R.B.~Vogelaar}
\author[Munchen]{F.~von~Feilitzsch}
\author[Krakow]{M.~Wojcik}
\author[Mainz]{M.~Wurm}
\author[Genova]{S.~Zavatarelli}
\author[Dresda]{K.~Zuber}
\author[Krakow]{G.~Zuzel}

\fntext[LNGSG]{Present affiliation: INFN Laboratori Nazionali del Gran Sasso, 67010 Assergi (AQ), Italy}
\fntext[ISPRA]{Present affiliation: Istituto Superiore per la Protezione e la Ricerca Ambientale, 00144 Roma, Italy}
\fntext[Bicocca]{Present affiliation: Dipartimento di Fisica, Università degli Studi e INFN Milano-Bicocca, 20126 Milano, Italy}
\fntext[IHEP]{Present affiliation: IHEP Institute of High Energy Physics, 100049 Beijing, China}
\fntext[GSI]{Present affiliation: GSI Helmholtzzentrum für Schwerionenforschung GmbH, 64291 Darmstadt, Germany}
\fntext[CALI]{Present affiliation: Department of Physics and Astronomy, University of California, Irvine, California, USA}
\fntext[MPI]{Present affiliation: Max-Planck-Institut für Kernphysik, 69117 Heidelberg, Germany}
\address{\bf{The Borexino Collaboration}}

\address[APC]{AstroParticule et Cosmologie, Universit\'e Paris Diderot, CNRS/IN2P3, CEA/IRFU, Observatoire de Paris, Sorbonne Paris Cit\'e, 75205 Paris Cedex 13, France}
\address[Dubna]{Joint Institute for Nuclear Research, 141980 Dubna, Russia}
\address[Genova]{Dipartimento di Fisica, Universit\`a degli Studi e INFN, 16146 Genova, Italy}
\address[Krakow]{M.~Smoluchowski Institute of Physics, Jagiellonian University, 30348 Krakow, Poland}
\address[Kiev]{Institute for Nuclear Research of NASU, 03028 Kyiv, Ukraine}
\address[Kurchatov]{National Research Centre Kurchatov Institute, 123182 Moscow, Russia}
\address[Kurchatovb]{ National Research Nuclear University MEPhI (Moscow Engineering Physics Institute), 115409 Moscow, Russia}
\address[LNGS]{INFN Laboratori Nazionali del Gran Sasso, 67010 Assergi (AQ), Italy}
\address[Milano]{Dipartimento di Fisica, Universit\`a degli Studi e INFN, 20133 Milano, Italy}
\address[Perugia]{Dipartimento di Chimica, Biologia e Biotecnologie, Universit\`a degli Studi e INFN, 06123 Perugia, Italy}
\address[Peters]{St. Petersburg Nuclear Physics Institute NRC Kurchatov Institute, 188350 Gatchina, Russia}
\address[Princeton]{Physics Department, Princeton University, Princeton, NJ 08544, USA}
\address[PrincetonChemEng]{Chemical Engineering Department, Princeton University, Princeton, NJ 08544, USA}
\address[UMass]{Amherst Center for Fundamental Interactions and Physics Department, University of Massachusetts, Amherst, MA 01003, USA}
\address[Virginia]{Physics Department, Virginia Polytechnic Institute and State University, Blacksburg, VA 24061, USA}
\address[Munchen]{Physik-Department, Technische Universit\"at  M\"unchen, 85748 Garching, Germany}
\address[Lomonosov]{Lomonosov Moscow State University Skobeltsyn Institute of Nuclear Physics, 119234 Moscow, Russia}
\address[GSSI]{Gran Sasso Science Institute, 67100 L'Aquila, Italy}
\address[Dresda]{Department of Physics, Technische Universit\"at Dresden, 01062 Dresden, Germany}
\address[Mainz]{Institute of Physics and Excellence Cluster PRISMA+, Johannes Gutenberg-Universit\"at Mainz, 55099 Mainz, Germany}
\address[Juelich]{Institut f\"ur Kernphysik, Forschungszentrum J\"ulich, 52425 J\"ulich, Germany}
\address[RWTH]{III. Physikalisches Institut B, RWTH Aachen University, 52062 Aachen, Germany}
\address[London]{Department of Physics, Royal Holloway University of London, Egham, Surrey,TW20 0EX, UK}
\address[Atomki]{Institute of Nuclear Research (Atomki), Debrecen, Hungary}
\address[Aquila]{Dipartimento di Scienze Fisiche e Chimiche, Universit\`a dell'Aquila, 67100 L'Aquila, Italy}


\begin{abstract}

In this paper, we report the first measurement of CNO solar neutrinos by Borexino that uses the Correlated Integrated Directionality (CID) method, exploiting the sub-dominant Cherenkov light in the liquid scintillator detector. The directional information of the solar origin of the neutrinos is preserved by the fast Cherenkov photons from the neutrino scattered electrons, and is used to discriminate between signal and background. 
The directional information is independent from the spectral information on which the previous CNO solar neutrino measurements by Borexino were based. While the CNO spectral analysis could only be applied on the Phase-III dataset, the directional analysis can use the complete Borexino data taking period from 2007 to 2021. The absence of CNO neutrinos has been rejected with $>$5$\sigma$ credible level using the Bayesian statistics. The directional CNO measurement is obtained without an external constraint on the $^{210}$Bi contamination of the liquid scintillator, which was applied in the spectral analysis approach. The final and the most precise CNO measurement of Borexino is then obtained by combining the new CID-based CNO result with an improved spectral fit of the Phase-III dataset. Including the statistical and the systematic errors, the extracted CNO interaction rate is $R\mathrm{(CNO)}=6.7^{+1.2}_{-0.8} \, \mathrm{cpd/100\, tonnes}$. Taking into account the neutrino flavor conversion, the resulting CNO neutrino flux at Earth is $\Phi_\mathrm{CNO}=6.7 ^{+1.2}_{-0.8}  \times 10^8 \, \mathrm{cm^{-2} s^{-1}}$, which is found to be in agreement with the high metallicity Standard Solar Models.
The results described in this work reinforce the role of the event directional information in large-scale liquid scintillator detectors and open up new avenues for the next-generation liquid scintillator or hybrid neutrino experiments. A particular relevance is expected for the latter detectors, which aim to combine the advantages from both Cherenkov-based and scintillation-based detection techniques.

   \end{abstract}

  
\maketitle

    \twocolumn 
    \tableofcontents

\section{\label{sec:Intro}Introduction}

Solar neutrinos are produced in the core of the Sun by nuclear reactions in which hydrogen is transformed into helium. The dominant sequence of reactions is the so-called $pp$ chain~\cite{Bahcall:1989ks,Vinyoles_New_Gen_SSM} which is responsible for most of the solar luminosity, while approximately 1\,\% of the solar energy is produced by the so-called Carbon-Nitrogen-Oxygen (CNO) cycle. Even though the CNO cycle plays only a marginal role in the solar fusion mechanisms, it is expected to take over the luminosity budget for main sequence stars more massive, older, and hotter than the Sun~\cite{Salaris}.
Solar neutrinos have proven to be a powerful tool to study the solar core ~\cite{Bx_nature_CNO,Bx_Nature_2018,superK_solar_neutrino_IV,SNOPlus_B8} and, at the same time, have been of paramount importance in shedding light on the neutrino oscillation phenomenon ~\cite{Cleveland:1998nv,SAGE:2009eeu,Kaether:2010ag,KamLAND:2004mhv,PhysRevLett.89.011301,Bellini:2011rx}.

One important open question concerning solar physics regards the metallicity of the Sun, that is, the abundance of elements with $Z>2$.
In fact, different analyses of spectroscopic data yield significantly different metallicity results, that can be grouped in two classes: the so-called High-Metallicity (HZ) \cite{HZ,HZ1} and Low-Metallicity (LZ) \cite{LZ,LZ1,LZ2} models.
The solar neutrino fluxes, in particular that from the CNO cycle reactions, can address this issue. Indeed, the SSM predictions of the CNO neutrino flux depend on the solar metallicity directly, via the abundances of C and N in the solar core, and indirectly, via its effect on the solar opacity and  temperature profile. 

Borexino delivered the first direct experimental proof of the existence of the CNO cycle in the Sun with a significance of $\sim$7\,$\sigma$, also providing a slight preference towards High-Metallicity models ~\cite{Bx_nature_CNO, Bx_improved_CNO}. 
This result was obtained with a multivariate analysis of the energy and radial distributions of selected events. To disentangle the CNO signal from the background, the multivariate fit requires an independent external constraint on the $pep$ neutrino rate and on the $^{210}$Bi rate; the latter is obtained by tagging $^{210}$Bi-$^{210}$Po coincidences in a temperature stabilized, layered scintillator fluid (see \cite{Bx_nature_CNO, Bx_improved_CNO} for more details).
For this reason, the CNO measurement has been performed only on approximately one third of the Borexino data, the so-called Phase-III.

In this paper, we present new results on CNO neutrinos obtained exploiting the "Correlated and Integrated Directionality" (CID) technique, which uses the directional information encoded in the Cherenkov light emitted alongside the scintillation, to separate the solar signal from non-solar backgrounds.
Borexino demonstrated the viability of this technique using $^7$Be solar neutrinos \cite{Bx_CID_long, Bx_CID_short}.
Here we apply the CID technique to the CNO analysis, obtaining two important results: we show that we can extract the evidence of solar CNO neutrinos on the entire Borexino dataset following an alternative approach with respect to the standard multivariate analysis and, consequently, without the help of the $^{210}$Bi constraint; we also show that by combining the information coming from the directionality with the standard multivariate analysis performed on Phase-III data we obtain an improved measurement of the CNO neutrino interaction rate.

The paper is structured as follows.
Section~\ref{sec:bxdet} describes the Borexino detector and summarizes the event reconstruction techniques. The CID analysis for the CNO neutrino measurement is illustrated in Sec.~\ref{sec:CID}, outlining the methods, reporting the results, and detailing the main sources of systematic uncertainties.
Finally, in Sec.~\ref{sec:MV} we show our best result on CNO neutrinos obtained combining the CID and the standard multivariate analysis.

\section{The Borexino experiment}

\label{sec:bxdet}

\begin{figure}[t]
    \centering
    \includegraphics[width=0.45\textwidth]{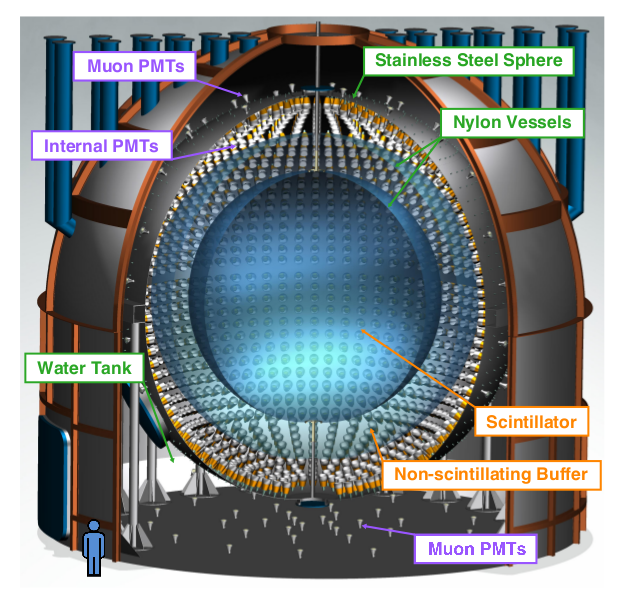}
    \caption{Scheme of the Borexino detector.}
    \label{fig:bx_det}
\end{figure}

Borexino was a liquid scintillator (LS) neutrino detector~\cite{bx_det} that ran until October 2021 with unprecedented radiopurity levels~\cite{Bx_long_paper,Bx_Nature_2018}, a necessary feature of its solar neutrino measurements.
The detector was located deep underground at the Laboratori Nazionali del Gran Sasso (LNGS) in Italy, with about \SI{3800}{m} water equivalent rock shielding suppressing the cosmic muon flux by a factor of $\sim$$10^6$. 

The detector layout is schematically shown in Fig.~\ref{fig:bx_det}. The Stainless Steel Sphere (SSS) with a 6.85\,m radius supported 2212 8-inch photomultiplier tubes (PMTs) and contained 280\,tonnes of pseudocumene (1,2,4-trimethylbenzene, PC) doped with 1.5\% of PPO (2,5-diphenyloxazole) wavelenght shifter, confined in a nylon inner vessel of 4.25\,m radius. The density of the scintillator was (0.878 $\pm$ 0.004)\,g\,cm$^{-3}$ with the electron density of (3.307 $\pm$ 0.015) $\times$ $10^{31}$ $e^-/100$\,tonnes. 
The PC-based buffer liquid in the region between the SSS and IV shielded the LS from external $\gamma$ radiation and neutrons.
The nylon Outer Vessel, that separated the buffer in two sub-volumes, prevented the inward diffusion of $^{222}$Rn.  The SSS itself is submerged in a domed, cylindrical tank filled with $\sim$1\,kton of ultra-pure water, equipped with 208 PMTs. The water tank provided shielding against external backgrounds and also served as an active Cherenkov veto for residual cosmic muons passing through the detector. 

Borexino detected solar neutrinos via their elastic scattering on electrons of the LS, a process sensitive, with different probability, to all neutrino flavors. Electrons, and charged particles in general, deposit their energy in the LS, excite its molecules and the resulting scintillation light is emitted isotropically. Using $n $$\approx$1.55 as scintillator index of refraction at 400\,nm wavelength, sub-dominant but directional Cherenkov light is emitted when the electron kinetic energy exceeds 0.165\,MeV. Cherenkov light is emitted over picosecond timescale while the fastest scintillation light component from the LS has an emission time constant at the nanosecond level. The fraction of light emitted as Cherenkov photons in Borexino was less than 0.5\% for 1 MeV recoiling electrons.

The effective total light yield was $\sim$500 photoelectrons per MeV of electron equivalent deposited energy, normalized to 2000 PMTs~\cite{Bx_long_paper}. 
The energy scale is intrinsically non-linear due to ionization quenching and the emission of Cherenkov radiation. The \textsc{Geant4} based Monte Carlo (MC) software \cite{Bx_monte_carlo} simulates all relevant physics processes. It is tuned using the data obtained during calibration campaigns with radioactive sources~\cite{Bx_calibration}. 
Distinct energy estimators have been defined, based on different ways of counting the number of detected photons~\cite{Bx_long_paper, Bx_CID_long}. 
The position reconstruction of each event is performed by using the time-of-flight corrected detection time of photons on hit PMT ~\cite{Bx_long_paper}. Particle identification is also possible in Borexino~\cite{Bx_long_paper}, in particular $\alpha/\beta$ discrimination~\cite{Bx_nature_CNO}, by exploiting different scintillation light emission time profiles.

The Borexino data-taking period is divided into three phases: Phase-I (May 2007-May 2010), Phase-II (December 2011-May 2016), and Phase-III (July 2016-October 2021). 
Phase-II started after the detector calibration~\cite{Bx_calibration} and an additional purification of the LS, that enabled a comprehensive measurement of the $pp$ chain solar neutrinos~\cite{Bx_Nature_2018}.
Phase-III is characterized by a thermally stable detector with greatly suppressed seasonal convective currents. This condition has made it possible to extract an upper limit constraint on the $^{210}$Bi contamination in the LS, and thus, to provide the first direct observation of solar CNO neutrinos~\cite{Bx_nature_CNO}. 

\section{Correlated and Integrated Directionality for CNO}
\label{sec:CID}

Cherenkov photons emitted by the electrons scattered in neutrino interactions retain information about the original direction of the incident neutrino. Typically, in water Cherenkov neutrino detectors, this information is accessed through an event-by-event direction reconstruction, as demonstrated by the measurements of $^{8}$B neutrinos, at energies larger than $\SI{3.5}{\mega\electronvolt}$~\cite{super-kamiokande_IV, SNOPlus_B8}.
Instead, the Borexino experiment has provided a proof-of-principle for the use of this Cherenkov hit information in a LS detector and at neutrino energies below $\SI{1}{\mega\electronvolt}$ through the so-called "Correlated and Integrated Directionality" (CID) technique. A detailed explanation of the method can be found in \cite{Bx_CID_short, Bx_CID_long}.

The CID method discriminates the signal originating in the Sun - due to solar neutrinos - from the background. Cherenkov light is sub-dominant in Borexino, but it is emitted almost instantaneously with respect to the slower scintillation light. Consequently, directional information is contained in the first hits of an event (after correcting for the time-of-flight of each photon). 
The CID analysis is based on the $\cos\alpha$ observable: for a given PMT hit in an event, $\alpha$ is the aperture angle between the Sun and the hit PMT at the reconstructed position of the event (see also Fig.~3 in ~\cite{Bx_CID_long}).
For background events, the $\cos\alpha$ distribution is nearly uniform regardless which hit is considered. For solar neutrino events, the $\cos\alpha$ distribution is flat for scintillation photons which are emitted isotropically, but has a characteristic non flat distribution peaked at $\cos\alpha\sim0.7$ for Cherenkov hits correlated with the position of the Sun.
Since we cannot distinguish Cherenkov and scintillation photons, in our previous work \cite{Bx_CID_short} we have used only the $1^\text{st}$ and $2^\text{nd}$ hits of each event, which have the largest probability of being Cherenkov hits.
In the new analysis presented in this paper we fully exploit the directional information contained in the first several hits.
This choice is supported by Monte Carlo simulations and sensitivity studies as discussed in Sec.~\ref{sec:Nth-hit}.

The solar neutrino signal is obtained by fitting the $\cos\alpha$ distributions of the selected first several hits, as a sum of signal and backgrounds contributions. The expected $\cos \alpha$ distributions for signal and background are obtained from Monte Carlo simulations.
As in Ref.~\cite{Bx_CID_long}, for each selected data event we simulate 200 MC events of solar neutrinos (represented by $^7$Be or $pep$ according to RoI, see below) and the same amount of the background events (represented by $^{210}$Bi). These events are simulated with the same astronomical time as the data event and with the position smeared around the reconstructed vertex.
From the fit we then obtain the total number of solar neutrinos $N_\nu$ detected in the RoI. The fit also includes two nuisance parameters. The effective Cherenkov group velocity correction $\text{gv}_\text{ch}$ nuisance parameter accounts for small differences in the relative hit time distribution between scintillation and Cherenkov hits in data relative to the MC. The second parameter is the event position mis-reconstruction in the initial electron direction $\Delta r_\text{dir}$, an indirect effect of the Cherenkov hits, where the reconstructed position is slightly biased towards early hit PMTs of the corresponding event. 
Here $\Delta r_\text{dir}$ is a free parameter of the fit, while $\text{gv}_\text{ch}$ is obtained independently and is constrained in the fit.

\begin{figure}[t]
    \centering
    \includegraphics[width=0.5\textwidth]
    {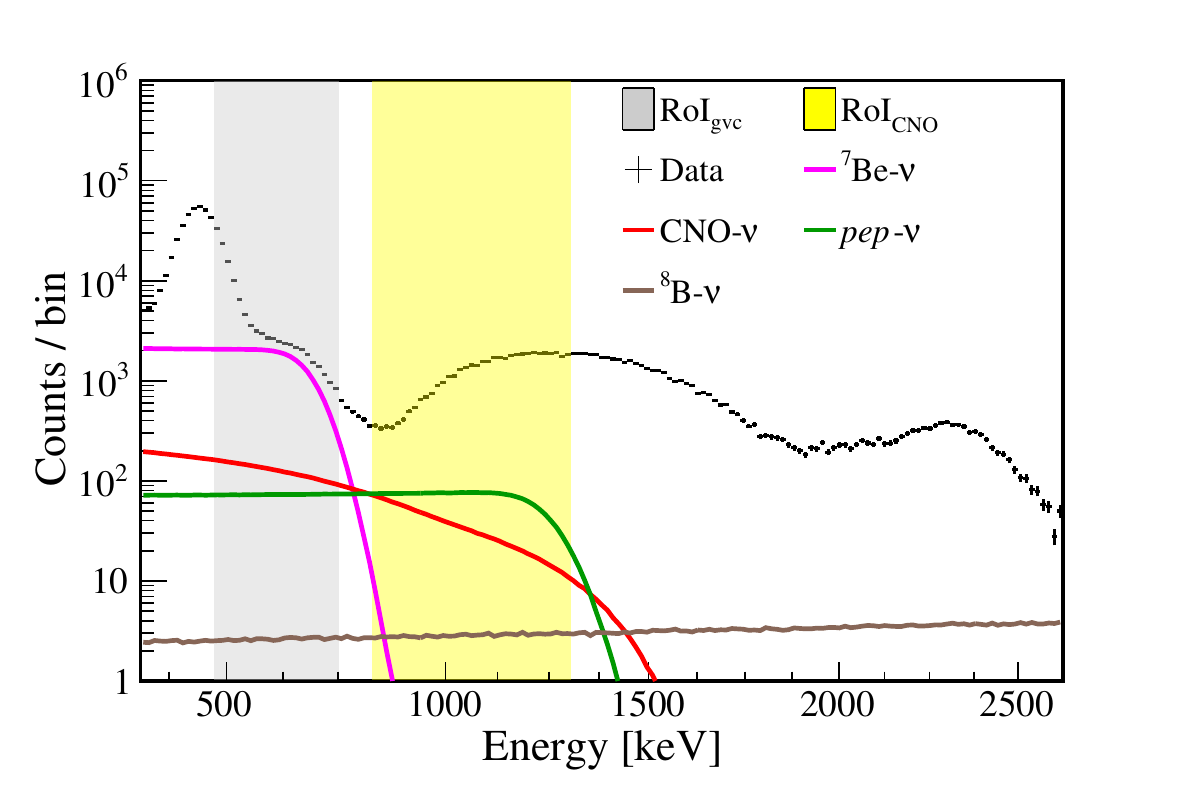}
    \caption{Illustration of the two RoIs used in the analysis on the energy spectrum of the Phase-II+III data in a fiducial volume of 2.95\,m radius. Monte Carlo PDFs of different solar neutrino components are scaled to high-metallicity SSM prediction~\cite{Vinyoles_New_Gen_SSM}. The grey band shows the $^{7}$Be-$\nu$ edge region used for the estimation of gv$_\text{ch}$ correction (RoI$_{\text{gvc}}$), while the CNO region used to measure the CNO-$\nu$ rate is shown in yellow band (RoI$_{\text{CNO}}$).}
    \label{fig:cid_ROI}
\end{figure}

Compared to the previous proof-of-principle analysis~\cite{Bx_CID_short, Bx_CID_long}, the current CID analysis has been improved in a variety of ways. 
The full detector live time can be used now thanks to a novel $\text{gv}_\text{ch}$ correction calibration.
In the previous publication, $\text{gv}_\text{ch}$ has been obtained using $^{40}$K $\gamma$ calibration data  (see \cite{Bx_CID_long}). In this work instead we calibrate $\text{gv}_\text{ch}$ by exploiting the $^7$Be solar neutrino events which allows us to extend the analysis to the full Borexino dataset, as explained in Sec.~\ref{sec:strategy_and_data}.
Additionally, indirect Cherenkov information from the systematic influence on the vertex reconstruction and consequently on the $\cos\alpha$ distribution was exploited by the inclusion in the analysis of later hits with negligible contribution of Cherenkov photons, see Sec.~\ref{sec:Nth-hit}. Technical details of the CID fitting procedure can be found in Sec.~\ref{sec: CID_fit}, while different systematic effects are discussed in Sec.~\ref{sec:cid_systematics}. The final CID results regarding the CNO measurement are reported in Sec.~\ref{subsec:CID_Results}.

\subsection{CID strategy for the CNO measurement with the full dataset}
\label{sec:strategy_and_data}

In the previous Borexino works~\cite{Bx_CID_short, Bx_CID_long}, the  calibration of the Cherenkov light group velocity $\text{gv}_\text{ch}$ has been performed using $\gamma$ sources deployed during the Borexino calibration campaign in 2009~\cite{Bx_CID_long}. The solar neutrino analysis was performed on the Phase-I dataset, that has been taken close in time to the source calibration of the detector ~\cite{Bx_calibration}. 
The $\text{gv}_\text{ch}$ found in this way was used to obtain the first measurement of $^7$Be solar neutrinos with the CID method \cite{Bx_CID_short}.
For the CID measurement of CNO in this paper, the entire Borexino dataset is used (from 2007 until 2021). Since the sub-nanosecond stability of the detector time response cannot be guaranteed for long periods, and no more calibrations have been performed after 2009, we developed a method to calibrate $\text{gv}_\text{ch}$ on the $^{7}$Be shoulder data.
This is done by using the same RoI as in ~\cite{Bx_CID_short, Bx_CID_long} (here called RoI$_{\text{gvc}}$ electron equivalent energy range of $\SI{0.5}{MeV} \lesssim T_e \lesssim \SI{0.8}{MeV}$) and performing the CID analysis where the $^7$Be is constrained to the Standard Model predictions ~\cite{Vinyoles_New_Gen_SSM}. 
The $\text{gv}_\text{ch}$ correction extracted in this way is then used in the CID analysis of the RoI$_{\text{CNO}}$, in which the CNO contribution is maximized, and which is fully independent from RoI$_{\text{gvc}}$. This step has been found to be justified according to MC studies, as the wavelength distribution of the detected Cherenkov photons produced by electrons from RoI$_{\text{gvc}}$ and RoI$_{\text{CNO}}$ is the same.
With this new strategy, the Cherenkov light $\text{gv}_\text{ch}$ can be calibrated on the same data-taking period as the one used for the CNO analysis. 
Two analyses have been performed in parallel for Phase-I (May 2007 to May 2010, 740.7\,days) and Phase-II+III (December 2011 to October 2021, 2888.0\,days).
The $\text{gv}_\text{ch}$ correction obtained for Phase-I can be compared to the one previously obtained from the $^{40}$K $\gamma$ source~\cite{Bx_CID_long}. Additionally, the analyses of the two independent data-sets allows for the investigation of any variation of the detector response over time.
The RoI$_{\text{gvc}}$ and the RoI$_{\text{CNO}}$ are shown for the Phase-II+III dataset in Fig.~\ref{fig:cid_ROI} for illustrative purposes.
The results on $\text{gv}_\text{ch}$ are provided in Sec.~\ref{subsub:gvcresults}.

In the final analysis, the Three-Fold-Coincidence algorithm~\cite{Bx_nature_CNO} is applied to the RoI$_{\text{CNO}}$ to suppress the cosmogenic $^{11}$C background, preserving the exposure with a signal survival fraction of $55.77\%\pm0.02\%$ for Phase-I and $63.97\%\pm0.02\%$ for Phase-II+III.
The radial ($R_\text{FV}$) and $T_e$ energy cuts of RoI$_{\text{CNO}}$ were optimized considering the expected number of solar neutrinos over the statistical uncertainty of the total number of events.
The optimized cuts are $R_\text{FV}<3.05 \,(2.95)$\,m and $0.85\,(0.85)\,\text{MeV} < T_e < 1.3\,(1.29)\,\text{MeV}$) for the Phase-I (Phase-II+III). In addition, all other cuts including the muon veto and data quality cuts have been applied as in Refs.~\cite{Bx_nature_CNO, Bx_improved_CNO}.
The overall exposures for the CID CNO analysis are $740.7\,\text{days} \times 104.3\,\text{tonnes}\times55.77\%$ for Phase-I and $2888.0\,\text{days} \times 94.4\,\text{tonnes} \times 63.97\%$ for Phase-II+III.
The total exposure of Phase-II+III (477.81\,years $\times$ tonnes) is about four times larger than that of Phase-I (118.04\,years $\times$ tonnes). 

\subsection{N$^\text{th}$-hit analysis approach}
\label{sec:Nth-hit}

As mentioned above, the CID analysis is performed on the first several early hits of ToF corrected hit times from each event in the RoI.
In this subsection we describe the optimization of the number of hits from each event to be used in the CID analysis. The procedure is based on the comparison of the MC-produced $\cos\alpha$ distributions of signal and background.

\begin{figure}[t]
    \centering
    \includegraphics[width=0.475\textwidth]
    {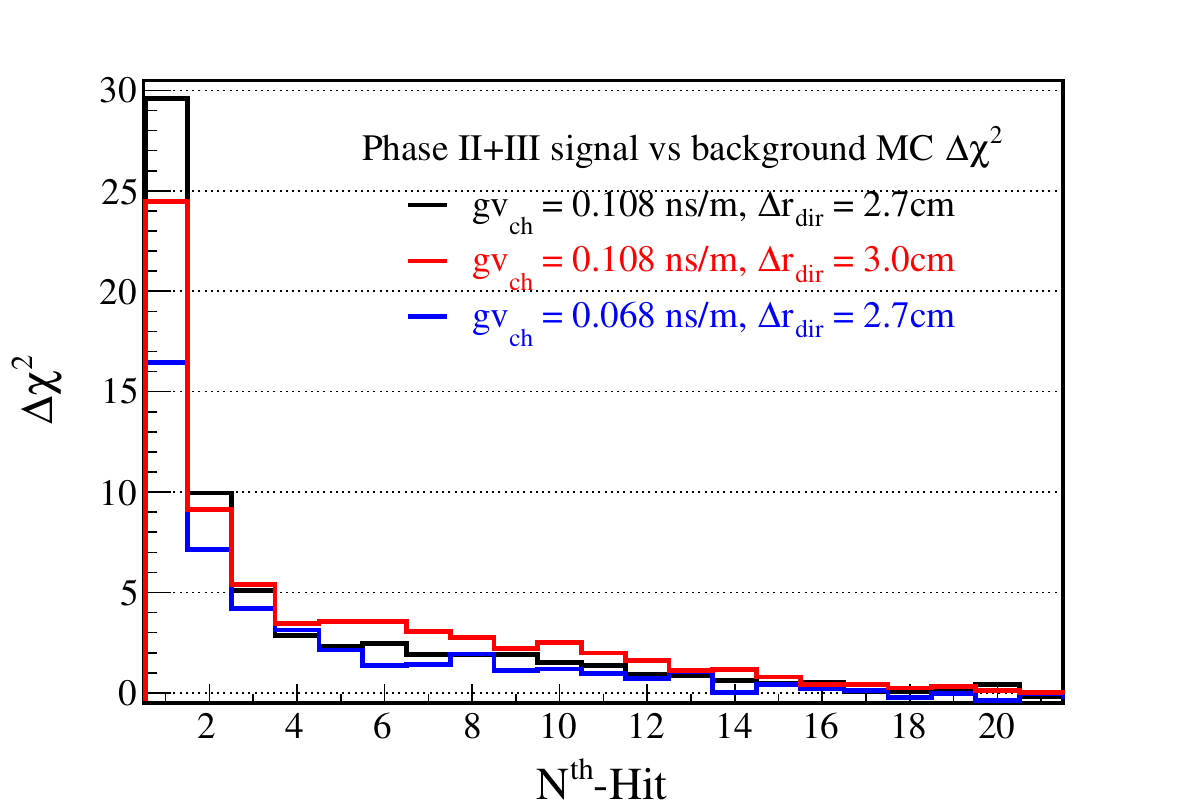}
    \caption{$\Delta\chi^2$ between the Phase-II+III RoI$_{\text{CNO}}$ neutrino signal and background MC $\cos\alpha$ distributions for different selections of nuisance parameters. $\Delta r_\text{dir} = 2.7\,\text{cm}$ corresponds to the nominal value observed in the neutrino MC.}
    \label{fig:cid_nth_hit_selection}
\end{figure}

First, PMT hits of each individual event are sorted according to their ToF-corrected hit times and are labeled in this order as "N$^\text{th}$-hit", with N = 1, 2, ... up to the total number of hits. Second, the $\cos\alpha$ distributions are constructed for each N$^\text{th}$-hit for both the signal and background MC. 
Third, for each N$^\text{th}$-hit $\cos\alpha$ distribution a number of 10,000 toy MC samples are simulated with the number of events observed as in the real data.
Next, we perform a direct signal to background comparison, based on a standard $\chi^2$-test.
Figure~\ref{fig:cid_nth_hit_selection} shows the resulting $\Delta\chi^2$ for Phase-II+III in the  RoI$_{\text{CNO}}$ averaged over the 10,000 toy datasets as a function of N$^\text{th}$-hit. 
A larger average $\Delta\chi^2$ corresponds to a greater difference between the MC signal and background and thus a larger expected sensitivity for the CID fit, independent of the true signal to background ratio.
While only the earliest $\sim$4 N$^\text{th}$-hits have a relevant, \textit{direct} contribution of Cherenkov hits to the $\cos\alpha$ distribution of the neutrino signal, later N$^\text{th}$-hits also contribute to the CID sensitivity due to the \textit{indirect} Cherenkov influence on $\Delta r_\text{dir}$.
A more in-depth explanation of the $\Delta r_\text{dir}$ effect is shown in Fig.~9 in ~\cite{Bx_CID_long}.

A possible impact of the $\text{gv}_\text{ch}$ and $\Delta r_\text{dir}$ nuisance parameters on the N$^\text{th}$-hit selection has been investigated and is presented in Fig.~\ref{fig:cid_nth_hit_selection}.
The first hits of the events provide the largest $\Delta\chi^2$ values thanks to the direct Cherenkov light. A decrease of $\text{gv}_\text{ch}$ is decreasing the group velocity of Cherenkov photons and thus their contribution at early hits. 
The impact of $\Delta r_\text{dir}$ can be seen for N$^\text{th}\text{-hit}>4$, where the contribution of direct Cherenkov hits becomes negligible relative to the scintillation hits, but the signal and background MC $\cos\alpha$ histograms are still different from each other ($\Delta\chi^2 > 0$).

In conclusion, the early hits selection for the CID analysis in both RoI$_{\text{gvc}}$ and RoI$_{\text{CNO}}$ is then performed from the first hit up to the $\text{N$^\text{th}$-hit(max)} = 15, 17$ for Phase-I and Phase-II+III, respectively. This is an optimization where all direct and indirect Cherenkov information is used, while at the same time this selection keeps the contribution from delayed scintillation photons, undergoing various optical process during the propagation through the detector, relatively small.

\begin{figure} [t]
    \centering
    \includegraphics[width=0.49\textwidth]{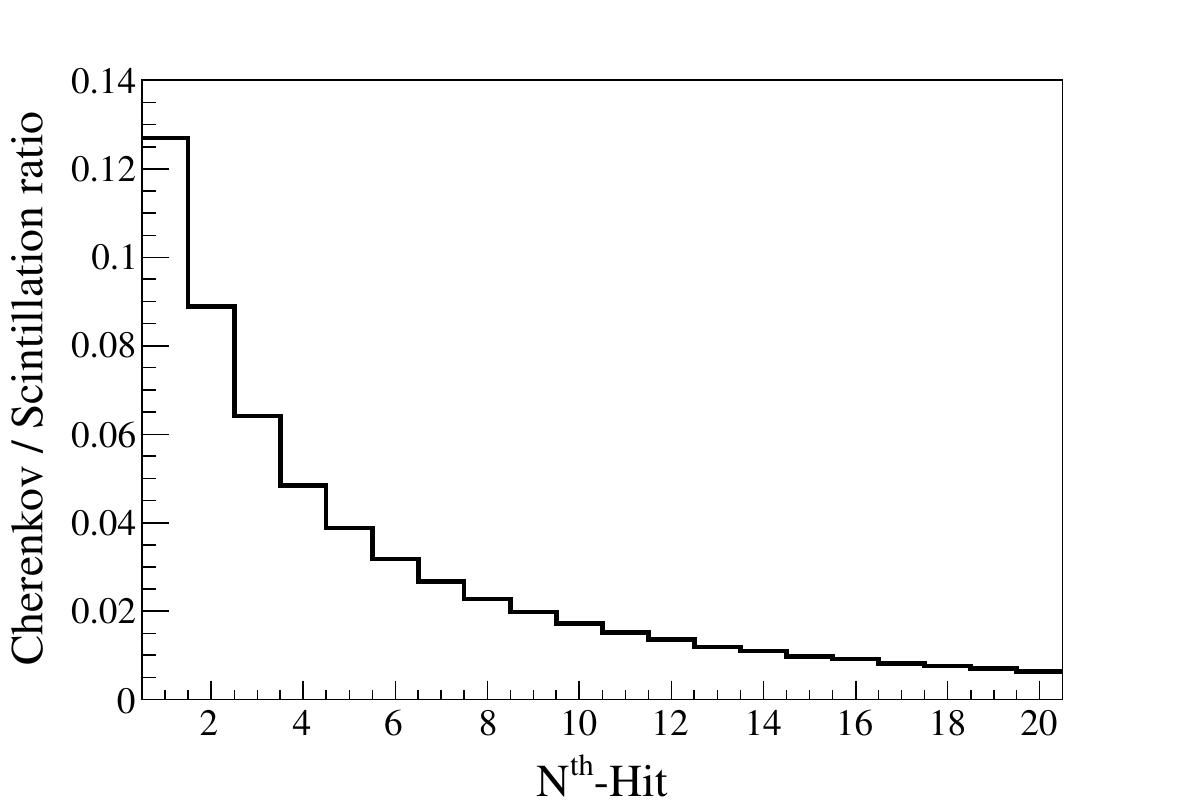}
    \caption{Cherenkov-to-scintillation PMT hit ratio as a function of the time-of-flight sorted N$^\text{th}$-hits for the $pep$ neutrino Monte Carlo of Phase II+III in the RoI$_{\text{CNO}}$ (0.85\,MeV -- 1.3\,MeV).}
    \label{fig:cheratio}
\end{figure}

The Cherenkov-to-scintillation photon ratio as a function of N$^\text{th}$-hit has also been checked explicitly, as is shown in Fig.~\ref{fig:cheratio} for RoI$_{\text{CNO}}$. As expected, it can be seen that the early  N$^\text{th}$-hits benefit from the largest Cherenkov-to-scintillation ratio of $\sim 13\%$ for the first hit. The overall total Cherenkov-to-scintillation ratio is small and found to be $0.475\%$ in the MC.

\subsection{CID fit procedure}
\label{sec: CID_fit}

The fitting strategy follows the procedure developed in our previous CID analysis~\cite{Bx_CID_long}.
The data $\cos\alpha$ distributions from the selected RoI, constructed for each N$^\text{th}$-hit from the first up to the N$^\text{th}$-hit(max),
are fitted simultaneously with the MC produced, expected $\cos\alpha$ distributions of the neutrino signal and background, where the signal $\cos\alpha$ distribution depends on $\text{gv}_\text{ch}$ and $\Delta r_\text{dir}$.
The nuisance parameter $\Delta r_\text{dir}$ cannot be calibrated in Borexino and is left free to vary without a dedicated pull term. The number of $\cos\alpha$ histogram bins used in the analyses is $i = 60$ for all energy regions and phases, as values of $i < 30$ reduce the expected CID sensitivity.

\subsubsection {Fit in the RoI$_{\text{gvc}}$}

The CID analysis in RoI$_{\text{gvc}}$ used for the $\text{gv}_\text{ch}$ calibration is based on the $\chi^{2}$-test:

\begin{align}
\begin{split}
    &\chi^{2}_{\text{gv}_\text{ch}}(N_{\nu}, \text{gv}_\text{ch}, \Delta r_\text{dir}) =\\
    &= \sum_{n=1}^\text{ N$^\text{th}$-hit(max)}\sum_{i=1}^{I} \left( \frac{ \left( \mathcal{N} \cdot M_{i}^{n} - D_{i}^{n} \right)^{2} }{ \mathcal{N} \cdot  M_{i}^{n} + \mathcal{N}^{2} \cdot M_{i}^{n} } \right) - 2 \ln\left(P(N_{\nu})\right),
\end{split}
\label{eq:cid_gvc_chi2_def}
\end{align}

\noindent where $D_{i}^{n}$ and  $M_{i}^{n}$ are the numbers of $\cos\alpha$ histogram entries at bin $i$ for a given N$^\text{th}$-hit $n$, for data and MC, respectively. The term $\mathcal{N}$ is the scaling factor between the MC and the data event statistics and the term "$\mathcal{N}^{2} \cdot M_{i}^{n}$" in the denominator takes into account the finite statistics of MC. The explicit dependence of the fit on $N_{\nu}$, $\text{gv}_\text{ch}$, and $\Delta r_\text{dir}$ can be expressed by decomposing the MC contribution to the one from the signal $S$ and the background $B$:
 
\begin{align}
M_{i}^{n} =\frac{N_\nu}{N_{\text{data}}} \cdot M_{\text{S},i}^{n}(\Delta r_{\text{dir}}, gv_{\text{ch}}) + (1 - \frac{N_\nu}{N_{\text{data}}}) \cdot M_{{\text{B}}, i}^{n}.
\label{eq:Mi}
\end{align}

\noindent The number of neutrino events $N_{\nu}$ and $\Delta r_\text{dir}$ are treated as nuisance parameters to produce the $\chi^{2}(\text{gv}_\text{ch})$ profile, where $N_{\nu}$ is constrained by the SSM expectation. For this, the neutrino prior probability distribution $P(N_{\nu})$ is given by the sum of the Gaussian probability distributions with mean and sigma from the high-metallicity~(HZ) SSM and low-metallicity~(LZ) SSM~\cite{Vinyoles_New_Gen_SSM} predictions on the number of $^{7}$Be+$pep$-$\nu$ in $^{7}$Be-$\nu$ shoulder region, which is then convoluted with a uniform distribution of CNO-$\nu$ between zero and the HZ-SSM CNO expectation + $5\sigma$. 
In this way, by leaving CNO reasonably free to vary, we avoid a potential correlation of the $\text{gv}_\text{ch}$ calibration and the subsequent measurement of the CNO-$\nu$ rate using this $\text{gv}_\text{ch}$ constraint.

\subsubsection {Fit in the RoI$_{\text{CNO}}$}

The $\chi^{2}$-test for the measurement of number of solar neutrinos ($N_{\nu}$) in RoI$_{\text{CNO}}$ is:

\begin{align}
\begin{split}
    &\chi^{2}_{\nu}(N_\nu, \text{gv}_\text{ch}, \Delta r_\text{dir}) =\\
    &= \sum_{n=1}^\text{ N$^\text{th}$-hit(max)}\sum_{i=1}^{I} \left( \frac{ \left( \mathcal{N} \cdot M_{i}^{n} - D_{i}^{n} \right)^{2} }{ \mathcal{N} \cdot  M_{i}^{n} + \mathcal{N}^{2} \cdot M_{i}^{n} } \right) + \Delta\chi_{\text{gv}_\text{ch}}^{2}\left(\text{gv}_\text{ch}\right)
\end{split}
\label{eq:cid_stt_chi2_def}
\end{align}

\noindent using $\text{gv}_\text{ch}$ and $\Delta r_\text{dir}$ as nuisance parameters. The $\text{gv}_\text{ch}$ parameter is now constrained by the previous calibration in RoI$_\text{gvc}$ through the pull term $\Delta\chi_{\text{gv}_\text{ch}}^{2}\left(\text{gv}_\text{ch}\right)$.

\subsection{Systematic uncertainties}
\label{sec:cid_systematics}

In this work we have performed a detailed evaluation of the systematic uncertainties. The quantitative evaluation of the systematic uncertainties on the CID analysis results are given in Sec.~\ref{subsec:CID_Results}.

It has been found that the choice of N$^\text{th}$-hit(max) and the histogram binning do not introduce any systematic uncertainty. Backgrounds different than $^{210}$Bi also contribute in the analysed energy intervals, but have been found to be indistinguishable in the CID analysis and do not contribute to the systematic uncertainty budget. Even if the external $\gamma$ events are not uniformly distributed in the FV, due to their attenuation in the LS, the difference in the $\cos\alpha$ distribution between these events and uniform background events is found to be safely negligible, given the statistics of the data.

The following effects have an impact on the final results:

 {\it PMT selection}
 
Some PMTs feature an intrinsic misbehaviour of their hit time distribution, if compared to all other PMTs. 
They are identified using an enriched sample of $^{11}$C events. In addition, a small number of PMTs feature inconsistency between the data and MC in the relative contribution to the first hits. The systematic effect has been evaluated by varying the selection of usable PMT.

{\it Relative PMT hit time correction}

It has been found by analyzing an enriched sample of $^{11}$C events, that the data PMTs have a small relative time offset between each other with an average value of 0.3\,ns. This time offset has been measured with an uncertainty of up to $\pm$0.1\,ns.
It is reasonable to correct this relative offset in data, as it does not exists in MC.
This makes it also necessary to propagate the uncertainty of the PMT time correction through the entire analysis chain, which introduces a systematic uncertainty.

{\it Influence of low number of signal events}

As described above, signal and background MC are produced on an event-by-event basis. This could  introduce an additional systematic uncertainty through the particular choice of the event positions and neutrino directions used for the production of the signal MC. This systematic uncertainty is estimated by producing a large number of the signal MC $\cos\alpha$ distributions corresponding to a random selection of the expected number of signal events from data in each phase. The data is then analyzed again with these reduced signal MC $\cos\alpha$ histograms. This effect contributes to the systematic uncertainty only in the RoI$_{\text{CNO}}$ for Phase-I.

{\it CNO-$\nu$ and $pep$-$\nu$ $\cos\alpha$ distributions}

The CNO-$\nu$ and $pep$-$\nu$ events show a significantly different energy distribution in the selected RoI. The expected Cherenkov to scintillation hits ratio for $pep$-$\nu$ events (0.475\%) is higher than for CNO events (0.469\%) due to their different energy distribution in RoI$_{\text{CNO}}$. Moreover, the angular distribution of recoiled electrons by CNO-$\nu$ and $pep$-$\nu$ is also different due to its dependence on energy distributions of the neutrino and recoiled electron. 
The final fit on the number of solar neutrinos is performed with the $pep$-$\nu$ MC and the systematic uncertainty is estimated by performing the CID analysis again with the CNO-$\nu$ MC.
The absolute difference between the two analyses is used conservatively as the systematic uncertainty. This systematic is negligible for the $\text{gv}_\text{ch}$ calibration, as CNO+$pep$ neutrinos are sub-dominant to the $^7$Be neutrinos.

{\it Constraint on non-CNO neutrinos}

The CID analysis is only sensitive to the measurement of the total number of solar neutrinos $N_\nu$ and cannot differentiate between different solar neutrino species.
Therefore, a measurement of the number of CNO neutrinos depends on the subtraction of the non-CNO neutrinos from $N_\nu$ obtained in the RoI$_{\text{CNO}}$. The number of $pep$ neutrinos is constrained by the SSM predictions~\cite{Vinyoles_New_Gen_SSM}, while $^{8}$B is constrained using the high precision flux measurement of Super-Kamiokande~\cite{superK_solar_neutrino_IV}. 

{\it Exposure }

In this category we cover systematic uncertainties on the determination of the fiducial mass and on the fraction of solar neutrinos in the RoI. These uncertainties are estimated using toy-MC studies, based on the results of the calibration campaign~\cite{Bx_calibration} on the performance of the position reconstruction and on the uncertainty on the energy scale described in~\cite{Bx_nature_CNO}, respectively. The uncertainty on the fiducial mass includes also the uncertainty on the scintillator density. Additionally, in the $\text{gv}_\text{ch}$ calibration using the constraint on the expected number of all solar neutrino events, we consider also the uncertainty on $\alpha / \beta$ discrimination applied to suppress $^{210}$Po $\alpha$ decays in the RoI$_{\text{gvc}}$.

\subsection{Results of the CID analysis}
\label{subsec:CID_Results}

\subsubsection {Effective $\text{gv}_\text{ch}$  calibration on the $^7$Be edge}
\label{subsub:gvcresults}

\begin{figure}[h]
    \centering
    \subfigure[]{\includegraphics[width=0.47\textwidth]{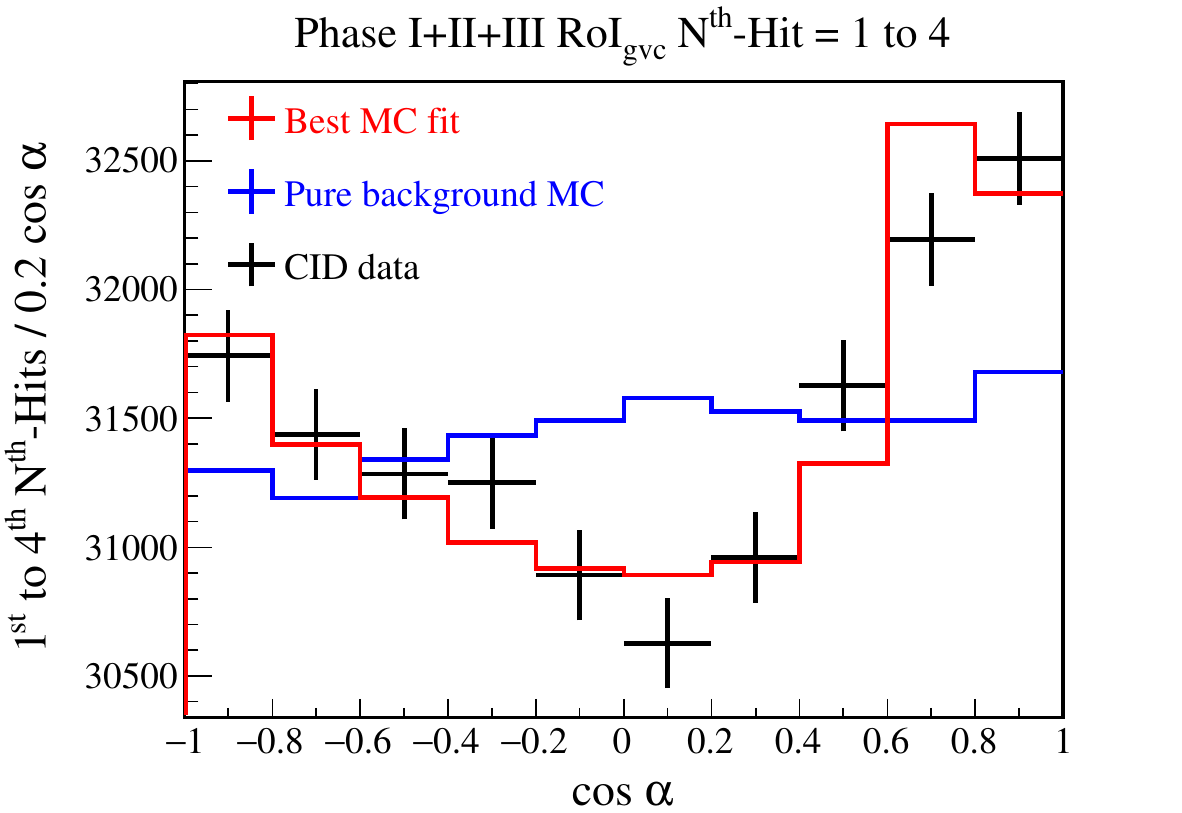}}
    \subfigure[]{\includegraphics[width=0.47\textwidth]{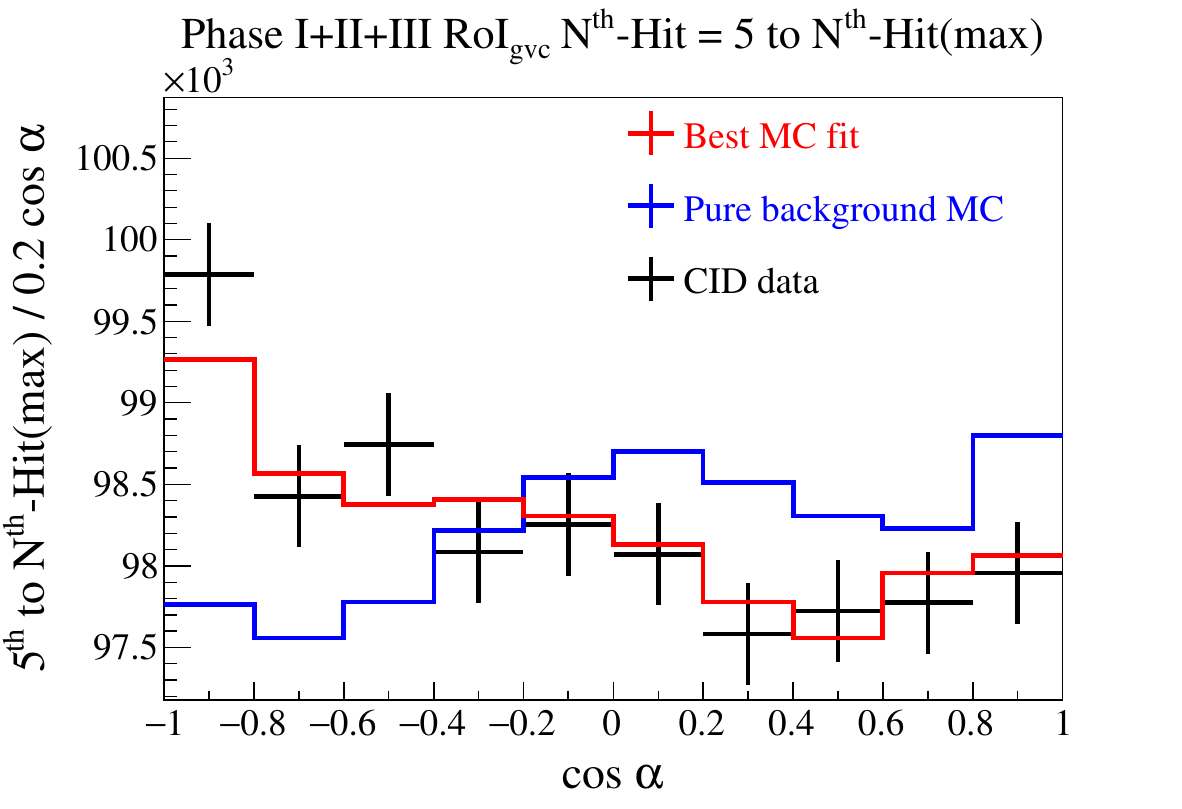 }}
    \caption{The CID data (black) and the best fit results (red) for the measurement of the $\text{gv}_\text{ch}$ parameter. 
    While the analysis is done separately for Phase-I and Phase-II+III, here the sum of Phase-I+II+III is shown for illustration purposes. There are in total 78632 events in the RoI$_{\text{gvc}}$.
    The best fit for the constrained number of neutrino events is $N_\nu = 50063$, while the best fit values for the parameter of interest are $\text{gv}_\text{ch} = 0.140\,\text{ns\,m}^{-1}$ for Phase-I and $\text{gv}_\text{ch} = 0.089\,\text{ns\,m}^{-1}$ for Phase-II+III.
    For comparison, the background MC (blue) scaled to the same total number of events is shown. (a) The sum of the first to fourth N$^\text{th}$-hits $\cos\alpha$ histograms shows the Cherenkov peak. (b) The sum of the fifth to the N$^\text{th}$-hit(max) $\cos\alpha$ histograms shows the effect the $\Delta r_\text{dir}$ parameter on the later hits.}
    \label{fig:bestgvc_cosalpha}
\end{figure}

The effective calibration of the Cherenkov light as a results of the CID analysis on the $^7$Be edge, using Eq.~\ref{eq:cid_gvc_chi2_def}, is $\text{gv}_\text{ch} = (0.140 \pm 0.029)\,\text{ns\,m}^{-1}$ for Phase-I and $\text{gv}_\text{ch} = (0.089 \pm 0.019)\,\text{ns\,m}^{-1}$ for Phase-II+III, including the systematic uncertainties summarized in Table~\ref{tab:cid_gvc_systeamtics}.
The compatibility between the data and the MC model, illustrated in Fig.~\ref{fig:bestgvc_cosalpha}, is good with $\chi^{2} / \text{ndf} = 874.9/897$, $p \text{ value} =0.70$ for Phase-I and $\chi^{2} / \text{ndf} = 1036.2/1017$, $p \text{ value} = 0.33$ for Phase-II+III. 
The $\chi^{2} / \text{ndf}$ and $p$ values have also been investigated for the individual N$^\text{th}$-hits $\cos\alpha$ histograms, with different binning choices to investigate the fit performance. For all cases, the best fit MC model $\cos\alpha$ distribution is always in agreement with the data.
Figure ~\ref{fig:bestgvc_cosalpha} shows an illustration of the best fit results (red) relative to a pure background hypothesis (blue).
For early hits, direct Cherenkov light causes the peak around $\cos\alpha\sim0.7$ and the influence of $\Delta r_\text{dir}$ induces the negative slope for $\cos\alpha < 0$.
For later N$^\text{th}$-hits the Cherenkov peak washes away, but the indirect impact of the Cherenkov hits on the position reconstruction bias $\Delta r_\text{dir}$ makes it still possible to distinguish between the neutrino signal and the background.
The non-flat background $\cos\alpha$ distribution originates from the live PMT non-isotropic distribution relative to the position distribution of the Sun around Borexino.
These $\text{gv}_\text{ch}$ values for Phase-I and Phase-II+III differ by less than $1.5\sigma$ and both are in agreement with the previous calibration performed at the end of Phase-I using a $^{40}$K $\gamma$ source: $\text{gv}_\text{ch} = (0.108 \pm 0.039)\,\text{ns\,m}^{-1}$  (Fig.\,13 in \cite{Bx_CID_long}).

\begin{table}[t]
\centering
\setlength\extrarowheight{2pt}
\begin{tabular}{c|cc}
\hline
\multicolumn{1}{c}{}                &  &  \\ [-13pt] \hline
Source of $\text{gv}_\text{ch}$ uncertainty & Phase-I & Phase-II+III \\ [2pt] \hline
PMT selection & 2.1\%     & 1.6\%       \\
PMT time corrections  & 3.7\%     & 2.1\%      \\
MLP event selection  & 1.0\%     & 1.0\%      \\
Fiducial mass & $\left(_{-1.2}^{+0.2}\right)\%$    & $\left(_{-1.2}^{+0.2}\right)\%$ \\
Fraction of neutrinos in RoI & 1.3\%    & 0.9\%      \\ \hline
\multicolumn{1}{c}{}                &  &  \\ [-13pt] \hline
\multicolumn{1}{c}{}                &  &  \\ [-10pt]
\end{tabular}
\caption{Systematic uncertainties of the $\text{gv}_\text{ch}$ measurement in the RoI$_{\text{gvc}}$, relative to the best fit value.}
\label{tab:cid_gvc_systeamtics}
\end{table}

\subsubsection {CNO measurement with CID} 
\label{subsub:CNOresults}

\begin{figure}[t]
    \centering
    \subfigure[]{\includegraphics[width=0.475\textwidth]{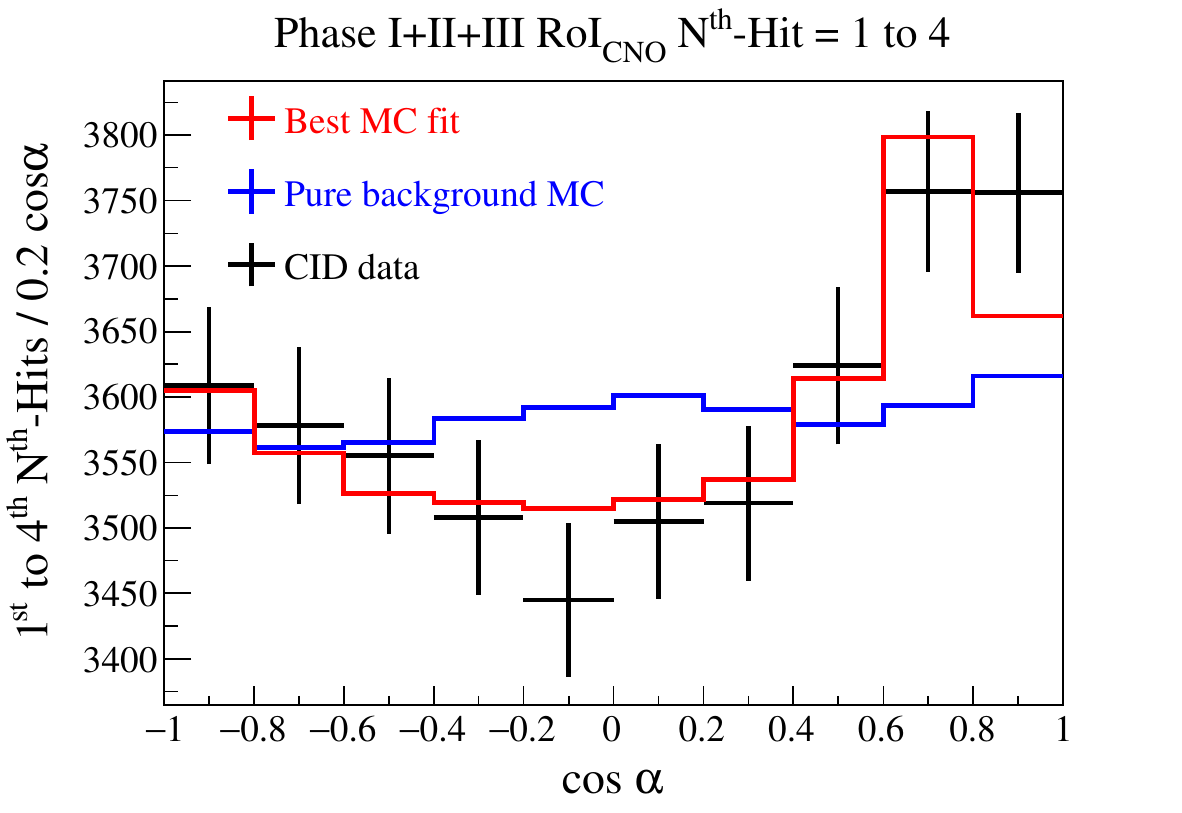}}
    \subfigure[]{\includegraphics[width=0.475\textwidth]{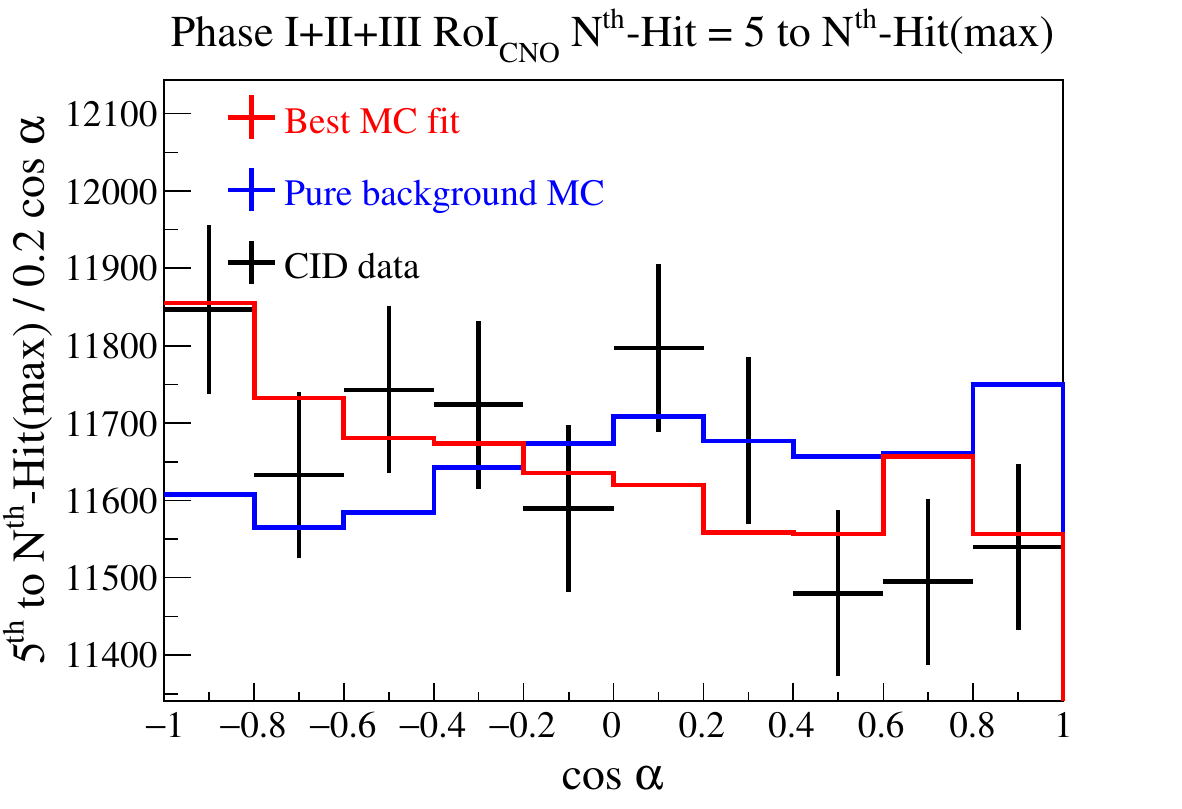 }}
    \caption{Illustration of the CID data (black) and the best fit results (red) summed for the Phase-I + Phase-II+III with a total of 8964 events in the RoI$_{\text{CNO}}$. The best fit of the total number of neutrino events is $N_\nu = 3519$ without any systematic correction. For comparison, the background MC (blue) scaled to the same total number of events is shown. (a) The sum of the first to fourth N$^\text{th}$-hits $\cos\alpha$ histograms shows the Cherenkov peak. (b) The sum of the fifth to the N$^\text{th}$-hit(max) $\cos\alpha$ histograms shows the effect the $\Delta r_\text{dir}$ parameter on these later hits.}
    \label{fig:bestfit_cosalpha}
\end{figure}

This section describes the results of the measurement of CNO solar neutrinos with CID using the full Borexino live time from May 2007 to October 2021. 
The $\text{gv}_\text{ch}$ values presented in Sec.~\ref{subsub:gvcresults} are used as independent pull terms in the Eq.~\ref{eq:cid_stt_chi2_def} for the fit in RoI$_{\text{CNO}}$ of their respective phases.
This takes into account the  potential systematic differences of the detector response between Phase-I and Phase-II+III. The resulting number of solar neutrino events $N_{\nu}$ in the RoI$_{\text{CNO}}$ can be converted into the number of CNO neutrinos detected in the same energy region after constraining the contributions from $pep$ and $^8$B neutrinos, but without any a-priori knowledge of the backgrounds. This number of CNO events can be further transformed into the measurement of the CNO-$\nu$ interaction rate in Borexino and the CNO flux at Earth. 


The best fit values for the number of solar neutrinos in RoI$_{\text{CNO}}$ are $N_\nu = 691_{-224}^{+235}\,\text{(stat)}$ for Phase-I and $N_\nu = 2828_{-494}^{+518}\,\text{(stat)}$ for Phase-II+III without inclusion of any systematic uncertainties or corrections. The compatibility betwwen the data and the MC model is good with $\chi^{2} / \text{ndf} = 884.8 / 897$, $p \text{ value} = 0.61$ for Phase-I and $\chi^{2} / \text{ndf} = 1000.7 / 1017$, $p \text{ value} = 0.64$ for Phase-II+III.
The MC model is able to reproduce the data $\cos\alpha$ distribution, which has also been investigated for the individual N$^\text{th}$-hits $\cos\alpha$ histograms.

Figure~\ref{fig:bestfit_cosalpha} illustrates the best fit results (red) relative to a pure background hypothesis (blue), in which the CID $\cos\alpha$ histograms of data (black) are shown for the sum of Phase-I + Phase-II+III, as well as for the sum of the early first to fourth N$^\text{th}$-hits (top) and the sum of the later N$^\text{th}$-hits from the fifth to $\mathrm{N^{th}-hit(max)}$ (bottom). The actual fit is performed on Phase-I and Phase-II+III independently.
The same observations made for Fig.~\ref{fig:bestgvc_cosalpha} hold also true for Fig.~\ref{fig:bestfit_cosalpha}, where the early hits show the Cherenkov peak and the later hits show the impact of $\Delta r_\text{dir}$.

\paragraph{Fit response bias correction}
\label{par:fit_response_bias}

Toy-MC analyses found that the fit of the number of solar neutrinos in RoI$_{\text{CNO}}$ shows a small systematic shift between the injected number of neutrinos and the best fit number of neutrinos, due to the correlation between the nuisance parameters ($\text{gv}_\text{ch}$, $\Delta r_\text{dir}$) and the relatively low total number of neutrino events.
This fit response bias is induced by the two nuisance parameters as they only impact the shape of the neutrino signal MC $\cos\alpha$ distribution but not that of background. We note that this effect was found to be negligible in RoI$_{\text{gvc}}$ due to the relative large number of signal events and the large signal to background ratio.

\begin{table}[t]
\centering
\setlength\extrarowheight{2pt}
\begin{tabular}{c|cc}
\hline
\multicolumn{1}{c}{}                &  &  \\ [-13pt] \hline
Source of uncertainty   & Phase-I & Phase-II+III \\ [2pt] \hline
\multicolumn{3}{c}{For N$_{\nu}$}   \\
\hline
PMT selection & 1.3\%     & 0.6\%       \\
PMT time corrections  & 4.2\%     & 2.4\%      \\
Low number of signal events & 2.2\%     &  --       \\
CNO-$\nu$ vs. $pep$-$\nu$ MC & 2.2\%     & 2.0\%      \\
\hline
\hline
\multicolumn{3}{c}{For N$_\text{CNO}$}     \\
[2pt] \hline
 &  &  \\ [-12pt]
$pep$+$^{8}$B-$\nu$ constraint & 4.6\%    & 1.8\%       \\ \hline
 \hline
\multicolumn{3}{c}{For R$_\text{CNO}$} \\ \hline
Fiducial mass                 & $\left(_{-1.2}^{+0.2}\right)\%$    & $\left(_{-1.2}^{+0.2}\right)\%$ \\
Fraction of CNO-$\nu$ in RoI & 1.4\%    & 1.4\%      \\ \hline
\multicolumn{1}{c}{}                &  &  \\ [-13pt] \hline
\multicolumn{1}{c}{}                &  &  \\ [-10pt]
\end{tabular}
\caption{Systematic uncertainties on the number of solar neutrino events $N_{\nu}$ in RoI$_{\text{CNO}}$, relative to the best fit value. The uncertainty from $pep$+$^{8}$B-$\nu$ constraint is relevant only for N$_\text{CNO}$. The last two rows are relevant only for the CNO-$\nu$ rate (R$_\text{CNO}$) calculation.}
\label{tab:cid_cno_systeamtics}
\end{table}

The value of the fit response bias in RoI$_{\text{CNO}}$ is estimated using the Bayesian posterior distribution of $N_{\nu}$~\cite{dagostini_bayes}, which is produced through a toy-MC rejection sampling, described in summary below.
The prior distribution for the number of neutrino events is chosen to be uniform between zero and the number of selected data events (2990 for Phase-I and 5974 for Phase-II+III), the prior distribution of $\Delta r_\text{dir}$ is also uniform, and the prior distribution of $\text{gv}_\text{ch}$ is given by the measurement at the $^7$Be-$\nu$ edge RoI$_{\text{gvc}}$ $\left(P\left(\text{gv}_\text{ch}\right) \propto \exp\left(-\frac{1}{2}\Delta\chi^2(\text{gv}_\text{ch})\right)\right)$.
The pseudo-data inputs $\left( N_{\nu}^{\text{sim}}, \text{gv}_\text{ch}^{\text{sim}}, \Delta r_\text{dir}^{\text{sim}} \right)$ are sampled from the MC signal and background $\cos\alpha$ distributions following these model parameter prior distributions.
The analysis is then performed in the same way as for the real data and results in best fit values of the pseudo-data $\left( N_{\nu}^{\text{fit}}, \text{gv}_\text{ch}^{\text{fit}}, \Delta r_\text{dir}^{\text{fit}} \right)$.
The real data result now defines a multivariate Gaussian distribution $\text{P}_\text{accept}(N_{\nu}, \text{gv}_\text{ch}, \Delta r_\text{dir})$ with a mean value given by its best fit values and with a standard deviation given by the systematic uncertainty of the PMT time corrections.
The sampled true values of the triplet  $\left( N_{\nu}^{\text{sim}}, \text{gv}_\text{ch}^{\text{sim}}, \Delta r_\text{dir}^{\text{sim}} \right)$ are then saved only with a probability of $\text{P}_\text{accept}(N_{\nu}^{\text{fit}}, \text{gv}_\text{ch}^{\text{fit}}, \Delta r_\text{dir}^{\text{fit}})$, given by the best fit result of the pseudo-data, otherwise they are rejected.
The resulting distributions of the true values for $\left( N_{\nu}, \text{gv}_\text{ch}, \Delta r_\text{dir} \right)$ then correspond to their Bayesian posterior distributions.

\begin{figure}[t]
    \centering
    \includegraphics[width=0.475\textwidth]{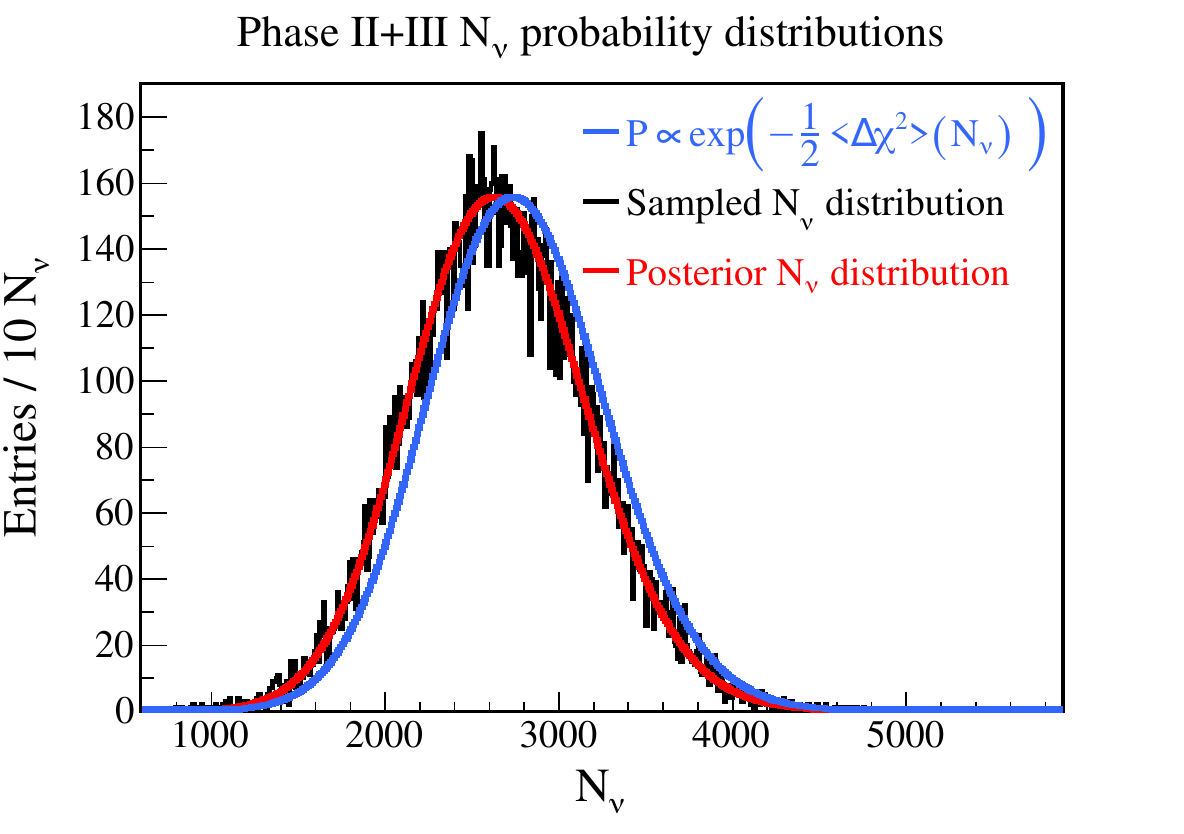}
    \caption{Illustration of the fit response bias for Phase II+III in the RoI$_{\text{CNO}}$. The data fit result, i.e. the likelihood $P(\nu)$ given by the $\Delta\chi^2$ profile of Eq.~\ref{eq:cid_stt_chi2_def} and averaged over 1000 fits with different PMT time offsets is shown in blue. The posterior distribution of 20k pseudo-data analyses, selected through rejection sampling, is shown in black. The red line corresponds to the posterior distribution that includes only the systematics from the PMT time alignment correction. }
    \label{fig:fit_response_bias}
\end{figure}

The fit response bias is illustrated in Fig.~\ref{fig:fit_response_bias} for Phase II+III. The likelihood distribution $P(N_\nu) \propto \exp \left(-\frac{1}{2} \Delta\chi^2(N_{\nu}\right)$ given by the $\chi^2$ fit of data with Eq.~\ref{eq:cid_stt_chi2_def} and averaged over the 1000 fits with different PMT time offsets is shown in blue. The black distribution is given by the simulation of 20k pseudo-data analyses, selected through the rejection sampling MC described above. The red distribution is produced by shifting $P(N_\nu)$ by a value of $\Delta N_\nu = -109\pm4$ and this distribution is well in agreement with the black rejection sampled distribution. It is therefore used as the posterior distribution of the CID analyses. We note that for the Phase-I the situation is similar and the shift is found to be $-50\pm4$ events.

\paragraph{Inclusion of systematics}
\label{par:fit_systematics}

The final result of the CID analysis for the number of solar neutrinos is given by the Bayesian posterior distribution of $N_\nu$, marginalized over the nuisance parameters and convoluted with the systematic uncertainties. The relevant systematic uncertainties are shown in Table~\ref{tab:cid_cno_systeamtics} and assumed to be normally distributed. 

\begin{figure}[t]
    \centering
    \includegraphics[width=0.475\textwidth]%
    {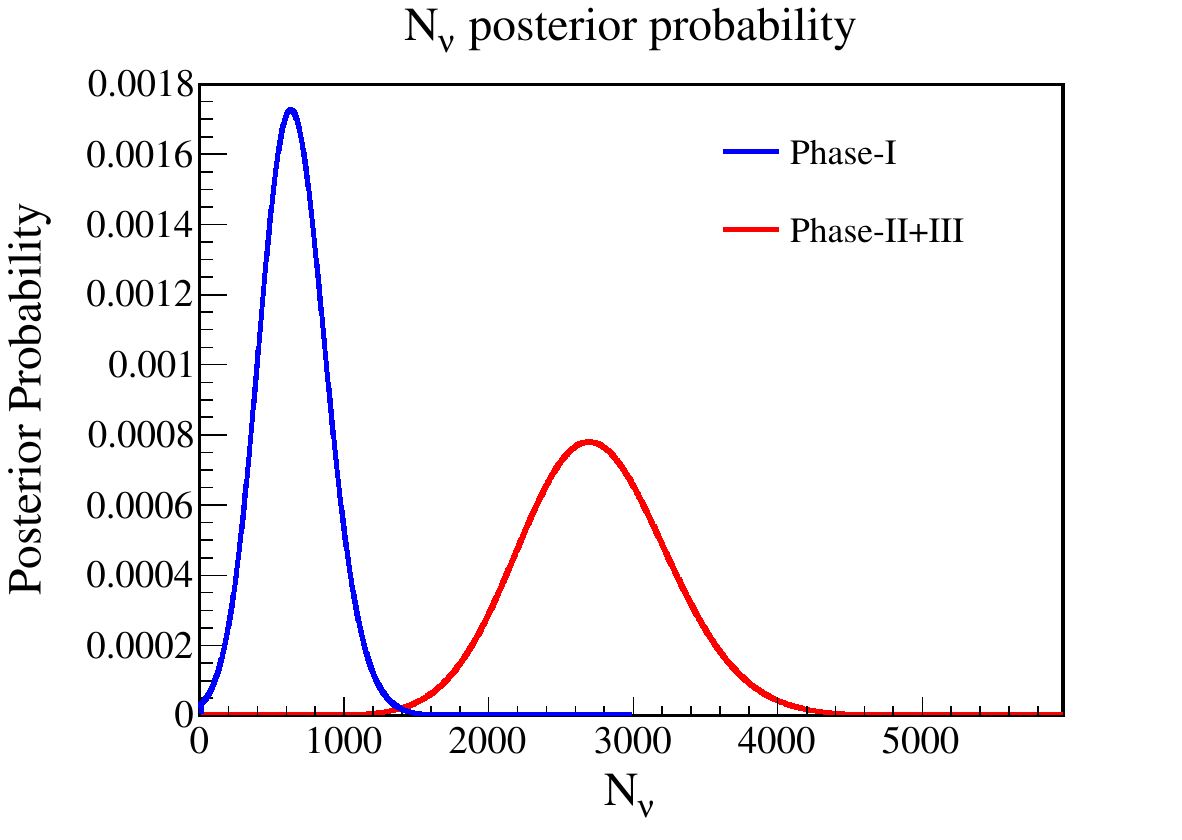}
    \caption{The CID measured posterior probability distributions for the number of solar neutrinos $N_{\nu}$ in the RoI$_{\text{CNO}}$ for Phase-I (blue) and Phase-II+III (red). All systematic effects are included.} 
    \label{fig:cid_numcnopep}
\end{figure}

The posterior distributions $P\left(N_\nu\right)$ in Phase-I and Phase-II+III including these systematics are shown in Fig.~\ref{fig:cid_numcnopep}. The resulting number of solar neutrinos detected in the RoI$_{\text{CNO}}$ is $N_\nu = 643_{-224}^{+235}\,\text{(stat)}_{-30}^{+37}\,\text{(sys)}$ for Phase-I and $N_\nu = 2719_{-494}^{+518}\,\text{(stat)}_{-83}^{+85}\,\text{(sys)}$ for Phase-II+III, including all systematics and correcting for the fit response bias. The quoted uncertainties are calculated from the posterior distributions using an $68\%$ equal-tailed credible interval (CI).
The one-sided zero neutrino hypothesis can be excluded with $P(N_\nu=0) = 2.8\times10^{-5}$ ($\sim4.2\sigma$) for Phase-I and $P(N_\nu = 0) = 6.4\times10^{-11}$ ($\sim6.5\sigma$) for Phase-II+III.

\paragraph{CID results on CNO}
\label{par:CIDCNO}

The interpretation of the CID results requires the correct treatment of the physical boundaries of the analysis, i.e. $0 \leq N_\nu \leq 2990$ (5974) for Phase-I (Phase-II+III), respectively.
This is done in a Bayesian interpretation, based on the posterior distribution $P\left(N_\nu\right)$ shown in Fig.~\ref{fig:cid_numcnopep}.

\begin{figure}[htb!]
    \centering
    \includegraphics[width=0.475\textwidth]{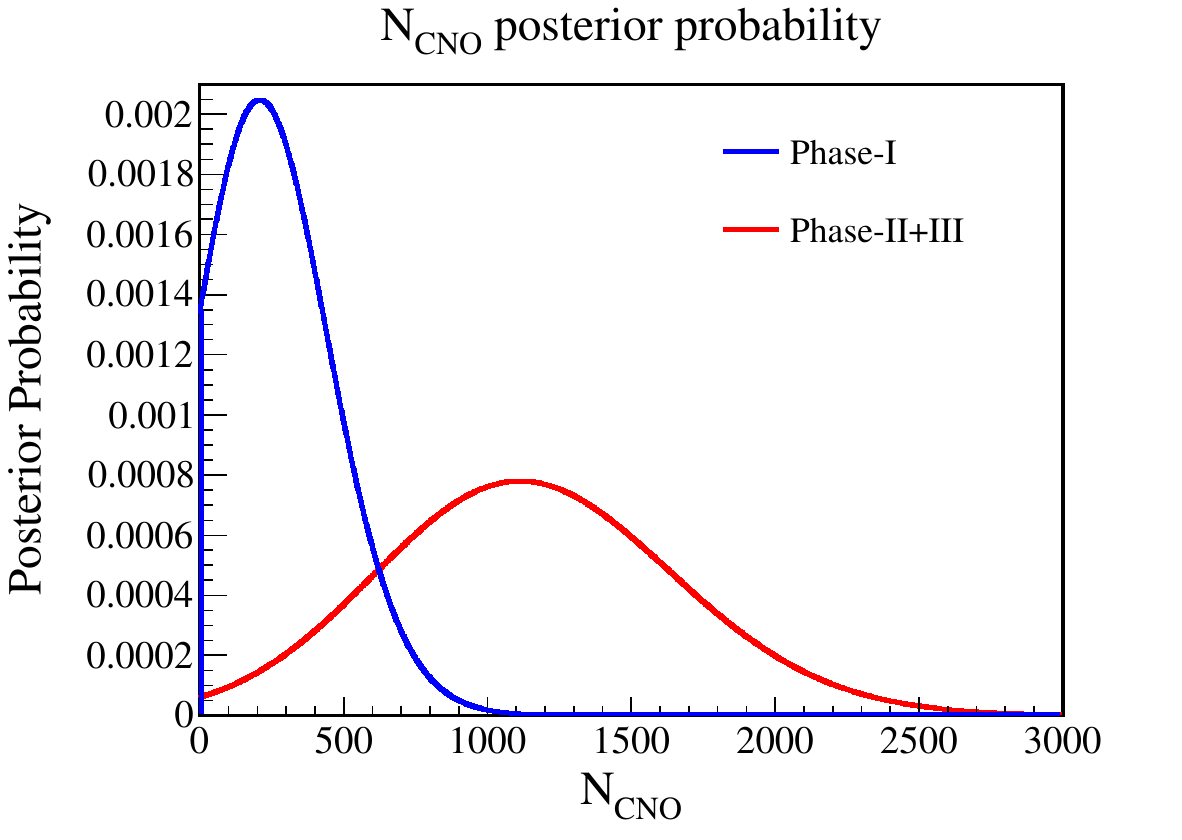}
    \caption{The CID measured posterior probability for the number of CNO-$\nu$ events after constraining $pep$ and $^8$B neutrinos for Phase-I (blue) and Phase-II+III (red). All systematic effects are included.
    }
    \label{fig:cid_numcno}
\end{figure}
\begin{figure}[htb!]
    \centering
    \includegraphics[width=0.475\textwidth]
    {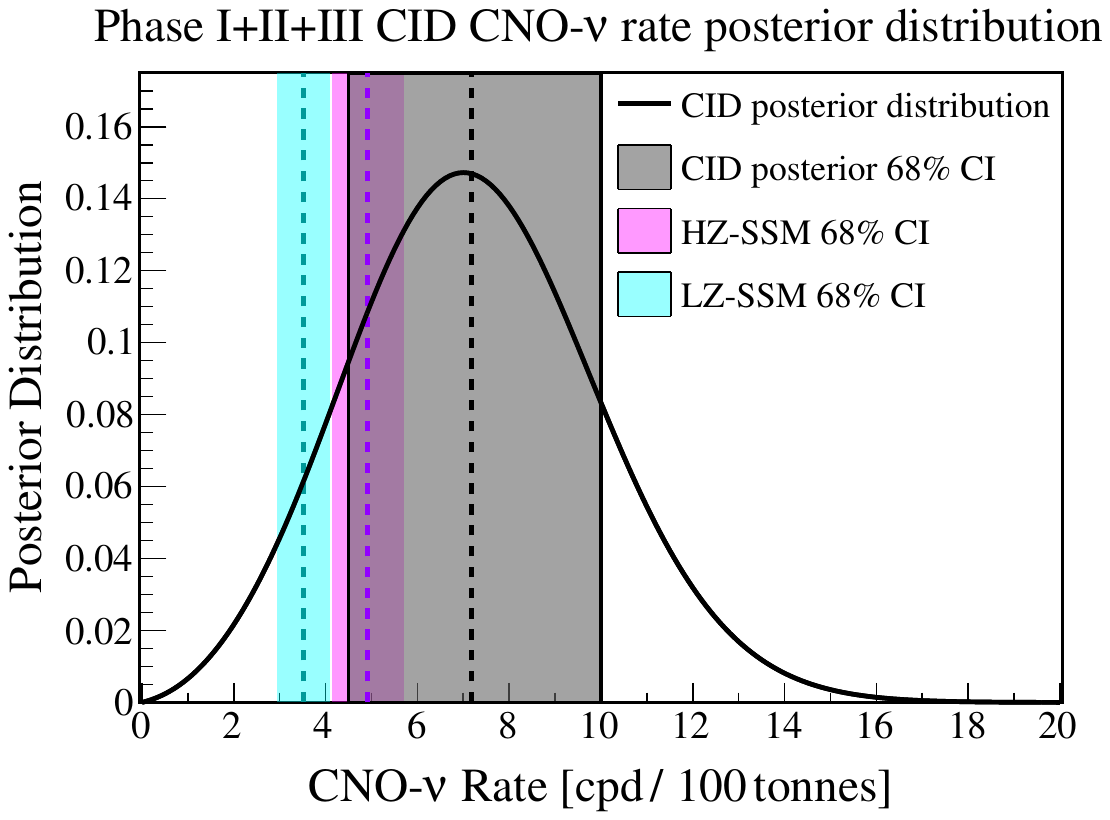}
    \caption{The combined CID Phase-I + Phase-II+III CNO-$\nu$ rate posterior distribution is shown in red. 
    The blue, violet and grey bands show the 68\% CI, for the low metallicity SSM B16-AGSS09met ($(3.52\pm0.52)\,\text{cpd\,/\,100\,tonnes}$), the high metallicity SSM B16-GS98 ($(4.92\pm0.78)\,\text{cpd\,/\,100\,tonnes}$) predictions~\cite{Vinyoles_New_Gen_SSM, Bx_CNO_sensitivity}, and the combined CID result, respectively. All systematic effects are included.}
    \label{fig:cid_combined_cnorate}
\end{figure}

Next, the distribution of the number of CNO-$\nu$ events is estimated by constraining the expected number of $pep$ and $^8$B neutrino events ($N_{pep+^8\text{B}}$) where the constraint on the number of $pep$ neutrinos uses the SSM predictions~\cite{Vinyoles_New_Gen_SSM} and $^{8}$B is constrained using the high precision flux measurement of Super-Kamiokande~\cite{superK_solar_neutrino_IV} including model uncertainties, the difference between HZ-SSM and LZ-SSM predictions, as well as the Borexino FV and energy systematic uncertainties from Table~\ref{tab:cid_cno_systeamtics}.
This is done through the convolution of the $N_\nu$ posterior distributions from Fig.~\ref{fig:cid_numcnopep} with the predicted $P(N_{pep+^8\text{B}})$ probability distribution: $P(N_\text{CNO}) = P(N_\nu)\ast P(-N_{pep+^8\text{B}})$.
The resulting $P(N_\text{CNO})$ posterior distributions are shown in Fig.~\ref{fig:cid_numcno}.
The CID measurement for the number of CNO-$\nu$ events is then $N_\text{CNO} = 270_{-169}^{+218}\,\text{(stat)}_{-25}^{+33}\,\text{(sys)}$ for Phase-I and $N_\text{CNO} = 1146_{-486}^{+518}\,\text{(stat)}_{-89}^{+92}\,\text{(sys)}$ for Phase-II+III, where the uncertainty corresponds to the equal-tail $68\%$ CI within the physical boundaries, including all systematics.

It has been observed that Phase-I and Phase-II+III do not show prohibitively different behavior for the full CID analysis-chain and the MC model is well able to reproduce the data $\cos\alpha$ histograms for both phase selections and for each selected energy region.
It is then reasonable to combine the conditionally independent results of Phase-I and Phase-II+III, through the convolution of both posterior distributions $P\left(N_\text{CNO}\right)^{\text{I+II+III}} = P\left(N_\text{CNO}\right)^{\text{I}} \ast P\left(N_\text{CNO}\right)^{\text{II+III}}$.
The probability that exactly zero CNO-$\nu$ events contribute to the measured data CID $\cos\alpha$ distribution is $P(N_\text{CNO}=0) = 1.35\times10^{-3}$ for Phase-I, $P(N_\text{CNO}=0) = 5.87\times10^{-5}$ for Phase-II+III, and $P(N_\text{CNO}=0) = 7.93\times10^{-8}$ for the combined result. This corresponds to a one-sided exclusion of the zero-CNO hypothesis at about $5.3\sigma$ credible level for the combination of Phase-I and Phase-II+III.

The CNO-$\nu$ rate probability density function is calculated from the measured posterior distribution of CNO-$\nu$ events, using the exposure of the respective phases. The effective exposure is given by the product of the fiducial mass, the detector live time, the TFC-exposure, the trigger efficiency, and the fraction of CNO-$\nu$ events within the selected energy region. 
The final CID result for the CNO-$\nu$ rate, using the full dataset of Phase-I + Phase-II+III, is $R_\text{CNO}^\text{CID} = 7.2\pm2.5\,\left(\text{stat}\right)\pm0.4\,\left(\text{sys}\right)\,_{-0.8}^{+1.1} \,\left(\text{nuisance}\right) \text{ cpd/100\,tonnes} = 7.2_{-2.7}^{+2.8} \, \text{ cpd/100\,tonnes}$
The quoted uncertainties now also show the systematic uncertainties from Table~\ref{tab:cid_cno_systeamtics} separately from the influence of the nuisance parameters $\text{gv}_\text{ch}$ and $\Delta r_\text{dir}$. The quoted statistical uncertainty corresponds to a hypothetical, perfect calibration of these CID nuisance parameters.
The results are summarized in Table~\ref{tab:cno_cno_final_result}.

\begin{table}[t]
\centering
\setlength\extrarowheight{4pt}
\begin{tabular}{c|cc}\hline
\multicolumn{1}{c}{}                &  &  \\ [-15pt] \hline
CID results        & $P(N_\text{CNO}=0)$ & $R_\text{CNO}$ $\left[ \frac{ \text{cpd} }{ \SI{100}{tonnes}} \right]$ \\[6pt] \hline
Phase-I   & $1.35\times10^{-3}$  & $6.4_{-4.1}^{+5.2}$\\[4pt]
Phase-II+III & $5.87\times10^{-5}$ & $7.3_{-3.2}^{+3.4}$ \\[4pt] \hline
Combined & $7.93\times10^{-8}$ & $7.2_{-2.7}^{+2.8}$ \\[4pt]  \hline
\multicolumn{1}{c}{}                &  &  \\ [-15pt] \hline
\multicolumn{1}{c}{}                &  &  \\ [-10pt]
\end{tabular}
\caption{CID CNO-$\nu$ results with systematic uncertainties.}
\label{tab:cno_cno_final_result}
\end{table}

These CID results are well in agreement with the HZ-SSM prediction of $(4.92\pm0.78) \text{ cpd/100\,tonnes}$ ($0.6\,\sigma$), while the LZ-SSM prediction $(3.52\pm0.52) \text{ cpd/100\,tonnes}$ ($1.1\,\sigma$) is 1.7 times less likely to be true, given the results of the Borexino CID analysis.

\section{Combined CID and multivariate analysis}
\label{sec:MV}

In this Section we combine the CID analysis with the standard multivariate fit of Phase-III to improve the result on CNO neutrinos.
This is done by including the posterior distributions of solar  neutrinos from the CID analysis, shown in Fig.~\ref{fig:cid_numcnopep}, in the multivariate analysis likelihood. By statistically subtracting the sub-dominant $^{8}$B neutrinos contribution and converting $N_\text{CNO+pep}$ to the corresponding interaction rate, it is possible to use these posterior distributions as external likelihood terms in the minimization routine. Following this procedure, two multiplicative pull terms  constraining the number of CNO and $pep$ neutrino events are used: the first one is related to the Phase-I ($\mathcal L^\mathrm{P-I}_\mathrm{CID}$), while the second one refers to Phase-II+III datasets ($\mathcal L^\mathrm{P-II+III}_\mathrm{CID}$).

The overall combined likelihood used for this analysis becomes:
\begin{equation}
     \mathcal L_\mathrm{MV +CID} = \mathcal L_\mathrm{MV} \cdot \mathcal L_\mathrm{pep} \cdot \mathcal L_\mathrm{^{210}Bi} \cdot  \mathcal L^\mathrm{P-I}_\mathrm{CID} \cdot  \mathcal L^\mathrm{P-II+III}_\mathrm{CID} 
     \label{eqn:MV_Likelihood_Full}
\end{equation}
where the first three terms correspond to an improved version of the standard multivariate analysis described in \cite{Bx_improved_CNO}.  This approach couples the one-dimensional Poisson likelihood for the TFC-Tagged dataset with a two-dimensional one (energy and radius) for the TFC-subtracted, to enhance the separation between signal and backgrounds.  We have improved the binning optimization and used an updated version of Monte Carlo.

The $pep$ neutrinos interaction rate is constrained with 1.4\% precision to the $2.74 \pm 0.04$ cpd/100 tonnes value, by combining the Standard Solar Model predictions~\cite{Vinyoles_New_Gen_SSM}, the most current flavor oscillation parameters set~\cite{Capozzi:2021fjo} and the solar neutrino data~\cite{Vescovi:2020wyz,Bergstrom:2016cbh}. 
This constraint is applied with the Gaussian pull term $\mathcal L_\mathrm{pep}$.
An upper limit on the $^{210}$Bi rate of ($10.8 \pm 1.0$ cpd/100 tonnes) is applied with the half Gaussian term $\mathcal L_\mathrm{^{210}Bi}$. This upper limit is obtained from the rate of the $^{210}$Bi daughter $^{210}$Po (see~\cite{Bx_nature_CNO,Bx_improved_CNO} for more details).

\subsection{Results} 
\label{subsec:MV_Results}

As in \cite{Bx_improved_CNO}, the energy RoI for the multivariate analysis is $\SI{0.32}{MeV} < T_e < \SI{2.64}{MeV}$ for electron recoil kinetic energy. The reconstructed energy spectrum scale is quantified in the $N_h$ estimator, representing total number of detected hits for a given event (see Sec.~\ref{sec:bxdet}). The dataset is the same one analyzed in ~\cite{Bx_improved_CNO}, in which the exposure amounts to 1431.6 days $\times$ 71.3 tonnes.

\begin{figure}[htb!]
    \centering
    \includegraphics[width=0.475\textwidth]
    {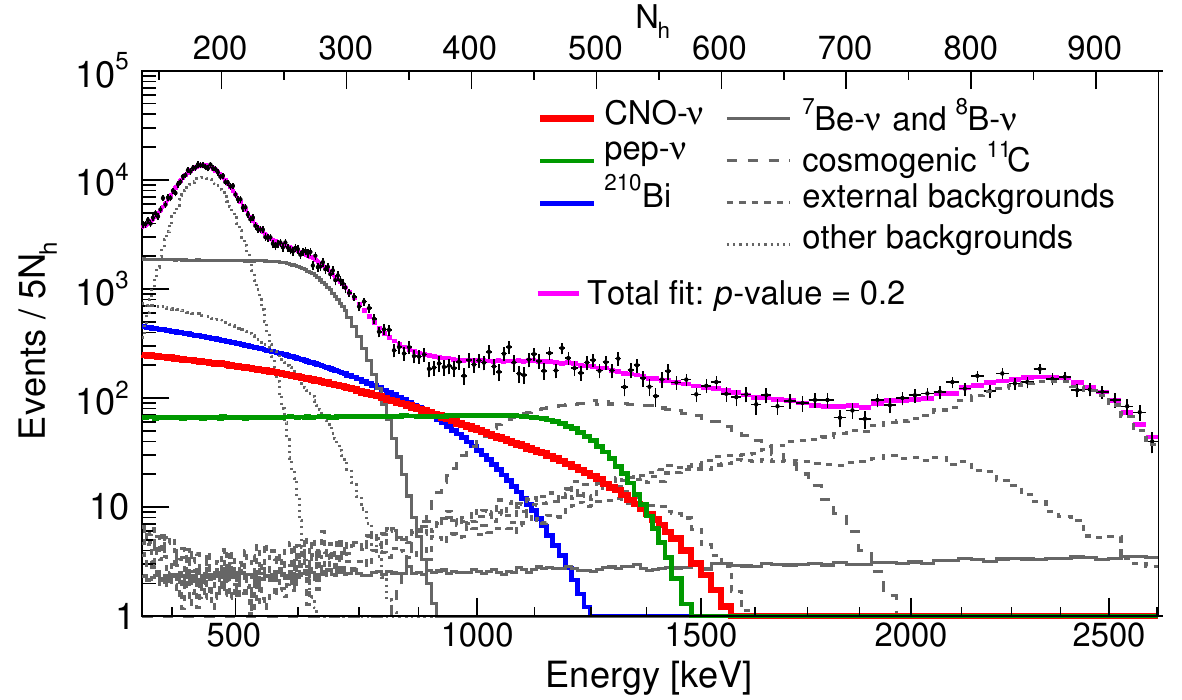}
    \includegraphics[width=0.475\textwidth]
    {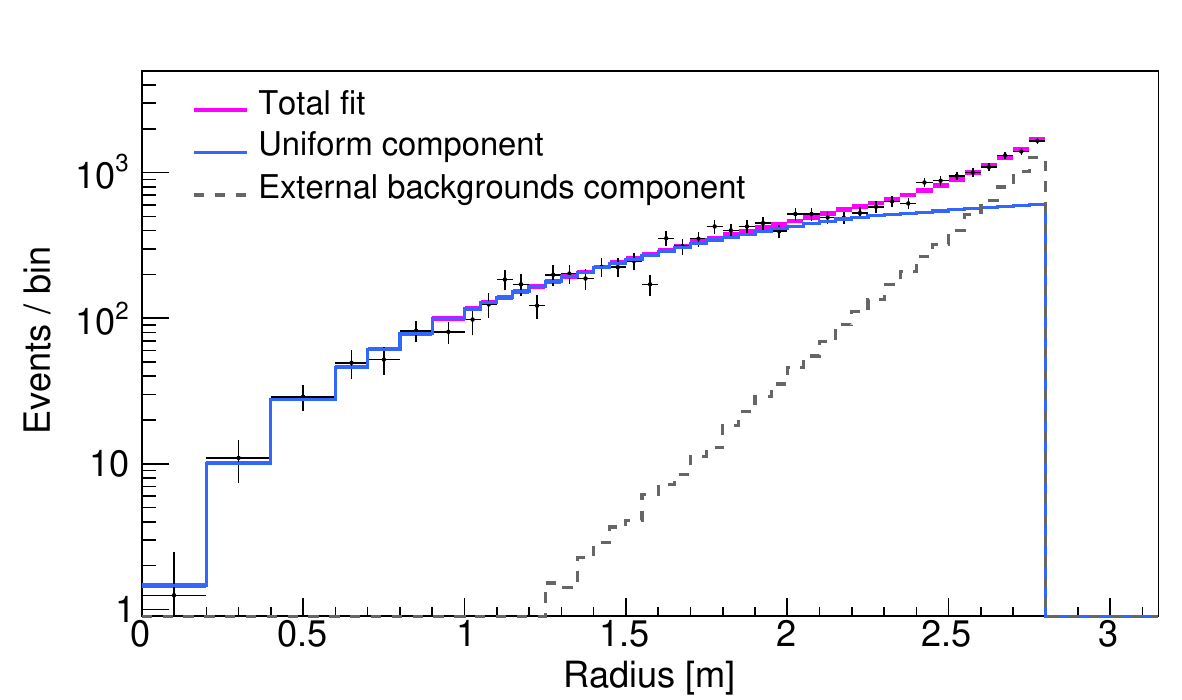}
    \caption{Multivariate fit results for the TFC-subtracted dataset, projected over the energy (top panel) and the radius (bottom panel) dimensions. For both projections, the sum of the individual components from the fit (magenta) is superimposed on the data (grey points). CNO neutrinos, $^{210}$Bi and pep neutrinos contributions are displayed in solid red, dashed blue and dotted green lines, respectively, while the other spectral components ($^7$Be and $^8$B neutrinos, other backgrounds) are shown in grey. The analysis has been performed using $N_h$ as energy estimator and the conversion to keV energy scale was performed only for the plotting purposes. The radial fit components, that are the uniform and the external backgrounds contributions, are shown in solid blue and dashed grey lines respectively.} 
    \label{fig:mv_spectralfit}
\end{figure}

Along with CNO solar neutrinos, the free parameters of the fit are divided into three categories: internal ($^{85}$Kr and $^{210}$Po) and external ($^{208}$Tl, $^{214}$Bi, and $^{40}$K) backgrounds, cosmogenic backgrounds ($^{11}$C, $^{6}$He, and $^{10}$C), and solar neutrinos ($^{7}$Be). Since $^{8}$B solar neutrinos exhibit a flat and marginal contribution, the corresponding interaction rate is fixed at high-metallicity expectations from Solar Standard Model. As discussed in Eq.~\ref{eqn:MV_Likelihood_Full}, the interaction rates of $pep$ neutrinos and $^{210}$Bi background are constrained with likelihood pull terms, and CID results reported in Sec.~\ref{subsec:CID_Results} are accounted for as additional external constraints. The result of the fit for the energy and radial projections is shown in Fig.~\ref{fig:mv_spectralfit}.

The multivariate fit returns an interaction rate of CNO neutrinos of $6.7^{+1.2}_{-0.7} \, \mathrm{cpd/100 \, tonnes}$ (statistical error only). The agreement between the model and data is quantified with a $p$ value of 0.2. 

To account for sources of systematic uncertainty, the same Monte Carlo method described in~\cite{Bx_nature_CNO,Bx_improved_CNO} has been adopted. In a nutshell, hundred thousands Monte Carlo pseudo-experiments were generated, including relevant effects able to introduce a systematic error, such as the energy response non linearity and non uniformity, the time variation of the scintillator light yield and the different theoretical models for the $^{210}$Bi spectral shape. The analysis is performed on these pseudo datasets assuming the standard response to study how this impacts the final result on CNO, yielding a total systematic uncertainty of $^{+0.31}_{-0.24} \, \mathrm{cpd/100 \, tonnes}$. Other sources of systematic error are included in the estimation of the upper limit on $^{210}$Bi contamination, as discussed more in detail in ~\cite{Bx_nature_CNO,Bx_improved_CNO}. 

\begin{figure}[htb!]
    \centering
    \includegraphics[width=0.475\textwidth]
    {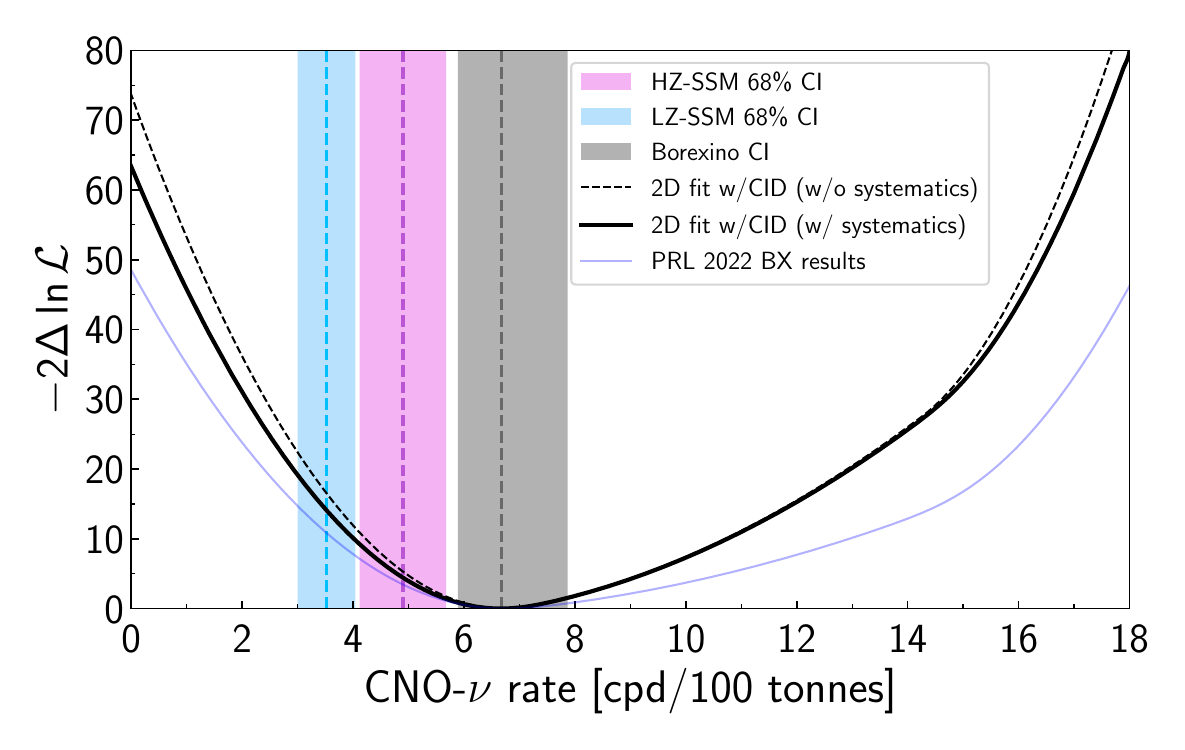}
    \caption{CNO-$\nu$ rate negative log-likelihood ($-2\Delta \ln \mathcal L$) profile obtained from the 2-dimensional multivariate spectral fit combined with the CID analysis constraint, with and without folding in the systematic uncertainties (black dashed and solid lines respectively). The blue line shows the same profile obtained in the previously published analysis without CID constraint ~\cite{Bx_improved_CNO}. The blue, violet, and gray vertical bands show 68\% confidence intervals (CI) for the low-metallicity SSM B16-AGSS09met ($3.52 \pm 0.52$ cpd/100\, tonnes) and the high-metallicity SSM B16-GS98 ($4.92 \pm 0.78$ cpd/100\,tonnes) predictions~\cite{Vinyoles_New_Gen_SSM,Bx_CNO_sensitivity}, and the new Borexino result including systematic uncertainty, respectively.} 
    \label{fig:results_dchi2}
\end{figure}

The negative log-likelihood profile as a function of the CNO rate is reported in Fig.~\ref{fig:results_dchi2}. The solid and dashed black lines show the results with and without systematic uncertainty, respectively. The result without CID constraint reported in ~\cite{Bx_improved_CNO} is included (blue line), for comparison. The improvement is clear especially for the upper value of the CNO rate.
The CNO interaction rate is  extracted from the 68\% quantile of the likelihood profile convoluted with the resulting systematic uncertainty, as $R^\mathrm{MV+CID}\mathrm{(CNO)}=6.7^{+1.2}_{-0.8} \, \mathrm{cpd/100 \, tonnes}$. The significance to the no-CNO hypothesis reaches about 8$\sigma$ C.L., while the resulting CNO flux at Earth is $\Phi(\mathrm{CNO})=6.7 ^{+1.2}_{-0.8} \times 10^8 \, \mathrm{cm^{-2} \, s^{-1}}$. 
Following the same procedure used in \cite{Bx_improved_CNO}, we use this result together with the $^8$B flux obtained from the global analysis of all solar data  to determine the abundance of C + N with respect to H in the Sun with an improved precision, for which we find $\rm{N_{\rm CN}} =5.81 ^{+1.22} _{-0.94} \times 10^{-4}$. This error includes both the statistical uncertainty due to the CNO measurement, and the systematic errors due to the additional contribution of the SSM inputs, to the ${^8}$B flux measurement, and to the $^{13}$N/$^{15}$O fluxes ratio. Similar to what was inferred from our previous publication, this result is in agreement with the High Metallicity measurements \cite{HZ, HZ1}, and features a 2$\sigma$ tension with Low Metallicity ones \cite{LZ,LZ1,LZ2}. 
Similarly, if we combine the new CNO result with the other Borexino results on $^7$Be and $^8$B in a frequentist hypothesis test based on a likelihood ratio test statistics we find that, assuming the HZ-SSM to be true, our data disfavours LZ-SSM at $3.2\sigma$ level.

\section{Conclusions}
\label{sec:conclusion}

In this work we have presented the results on CNO solar neutrinos obtained using the "Correlated and Integrated Directionality" (CID) technique.

We have shown that  the CID technique can be used to extract the CNO signal without any a priori assumptions on the backgrounds, in particular that of $^{210}$Bi. 
The Phase-I (May 2007 to May 2010, 740.7\,days) and Phase-II+III (December 2011 to October 2021, 2888.0\,days) datasets have been analyzed independently to investigate possible variations of the detector response over time. By adopting the Bayesian statistics, we have combined the conditionally independent results of Phase-I and Phase-II+III: the resulting CNO rate obtained with CID only is $7.2_{-2.7}^{+2.8} \, \text{cpd/100 \, tonnes}$. The no-CNO hypothesis including the $pep$ constraint only is rejected at 5.3$\sigma$ level. This result, albeit less precise than the one published by Borexino using the standard multivariate analysis, is the first obtained without the application of a $^{210}$Bi constraint.

We have also obtained an improved CNO solar neutrino result by combining the standard multivariate analysis with the CID technique.
The CID technique helps in separating the solar signal from non solar backgrounds, improving the significance and precision of the CNO measurement with respect to the result previously published by Borexino. 
The resulting CNO interaction rate is $6.7^{+1.2}_{-0.8} \, \mathrm{cpd/100\,tonnes}$ and the significance against the absence of a CNO signal, considered as the null hypothesis, is about 8$\sigma$.
The C+N abundance with respect to H is calculated from this result following the procedure adopted in \cite{Bx_improved_CNO} and is found to be $\rm{N_{\rm CN}} =5.81 ^{+1.22} _{-0.94} \times 10^{-4}$, compatible with the SSM-HZ metallicity measurements.

In conclusion, we have shown that the directional information of the Cherenkov radiation can be effectively combined with the spectral information coming from scintillation, for solar neutrino studies.
This combined detection approach provides a measurement that is more powerful than the individual methods on their own.
The sensitivity of the CID method could be significantly improved in future liquid scintillator-based detectors by optimizing the Cherenkov-to-scintillation ratio and by performing dedicated calibrations campaigns.

{\it Acknowledgments:}
We acknowledge the generous hospitality and support of the Laboratori Nazionali del Gran Sasso (Italy). The Borexino program is made possible by funding from Istituto Nazionale di Fisica Nucleare (INFN) (Italy), National Science Foundation (NSF) (USA), Deutsche Forschungsgemeinschaft (DFG), Cluster of Excellence PRISMA+ (Project ID 39083149), and recruitment initiative of Helmholtz-Gemeinschaft (HGF) (Germany), Russian Foundation for Basic Research (RFBR) (Grants No. 19-02-00097A), Russian Science Foundation (RSF) (Grant No. 21-12-00063) and Ministry of Science and Higher Education of the Russian Federation (Project FSWU-2023-0073) (Russia), and Narodowe Centrum Nauki (NCN) (Grant No. UMO 2017/26/M/ST2/00915) (Poland). We gratefully acknowledge the computing services of Bologna INFN-CNAF data centre and U-Lite Computing Center and Network Service at LNGS (Italy).

\bibliographystyle{apsrev4-1}
\bibliography{main.bib}

\begin{thebibliography}{31}%
\makeatletter
\providecommand \@ifxundefined [1]{%
 \@ifx{#1\undefined}
}%
\providecommand \@ifnum [1]{%
 \ifnum #1\expandafter \@firstoftwo
 \else \expandafter \@secondoftwo
 \fi
}%
\providecommand \@ifx [1]{%
 \ifx #1\expandafter \@firstoftwo
 \else \expandafter \@secondoftwo
 \fi
}%
\providecommand \natexlab [1]{#1}%
\providecommand \enquote  [1]{``#1''}%
\providecommand \bibnamefont  [1]{#1}%
\providecommand \bibfnamefont [1]{#1}%
\providecommand \citenamefont [1]{#1}%
\providecommand \href@noop [0]{\@secondoftwo}%
\providecommand \href [0]{\begingroup \@sanitize@url \@href}%
\providecommand \@href[1]{\@@startlink{#1}\@@href}%
\providecommand \@@href[1]{\endgroup#1\@@endlink}%
\providecommand \@sanitize@url [0]{\catcode `\\12\catcode `\$12\catcode
  `\&12\catcode `\#12\catcode `\^12\catcode `\_12\catcode `\%12\relax}%
\providecommand \@@startlink[1]{}%
\providecommand \@@endlink[0]{}%
\providecommand \url  [0]{\begingroup\@sanitize@url \@url }%
\providecommand \@url [1]{\endgroup\@href {#1}{\urlprefix }}%
\providecommand \urlprefix  [0]{URL }%
\providecommand \Eprint [0]{\href }%
\providecommand \doibase [0]{http://dx.doi.org/}%
\providecommand \selectlanguage [0]{\@gobble}%
\providecommand \bibinfo  [0]{\@secondoftwo}%
\providecommand \bibfield  [0]{\@secondoftwo}%
\providecommand \translation [1]{[#1]}%
\providecommand \BibitemOpen [0]{}%
\providecommand \bibitemStop [0]{}%
\providecommand \bibitemNoStop [0]{.\EOS\space}%
\providecommand \EOS [0]{\spacefactor3000\relax}%
\providecommand \BibitemShut  [1]{\csname bibitem#1\endcsname}%
\let\auto@bib@innerbib\@empty
\bibitem [{\citenamefont {Bahcall}(1989)}]{Bahcall:1989ks}%
  \BibitemOpen
  \bibfield  {author} {\bibinfo {author} {\bibfnamefont {J.~N.}\ \bibnamefont
  {Bahcall}},\ }\href@noop {} {\emph {\bibinfo {title} {{Neutrino
  Astrophysics}}}}\ (\bibinfo  {publisher} {Cambridge University Press},\
  \bibinfo {year} {1989})\BibitemShut {NoStop}%
\bibitem [{\citenamefont {Vinyoles}\ \emph {et~al.}(2017)\citenamefont
  {Vinyoles} \emph {et~al.}}]{Vinyoles_New_Gen_SSM}%
  \BibitemOpen
  \bibfield  {author} {\bibinfo {author} {\bibfnamefont {N.}~\bibnamefont
  {Vinyoles}} \emph {et~al.},\ }\href {\doibase 10.3847/1538-4357/835/2/202}
  {\bibfield  {journal} {\bibinfo  {journal} {Astrophysical Journal}\ }\textbf
  {\bibinfo {volume} {835}},\ \bibinfo {pages} {202} (\bibinfo {year}
  {2017})}\BibitemShut {NoStop}%
\bibitem [{\citenamefont {{Salaris}}\ and\ \citenamefont
  {{Cassisi}}(2005)}]{Salaris}%
  \BibitemOpen
  \bibfield  {author} {\bibinfo {author} {\bibfnamefont {M.}~\bibnamefont
  {{Salaris}}}\ and\ \bibinfo {author} {\bibfnamefont {S.}~\bibnamefont
  {{Cassisi}}},\ }\href@noop {} {\emph {\bibinfo {title} {{Evolution of Stars
  and Stellar Populations}}}}\ (\bibinfo  {publisher} {John Wiley \& Sons,
  Ltd},\ \bibinfo {year} {2005})\BibitemShut {NoStop}%
\bibitem [{\citenamefont {Agostini}\ \emph
  {et~al.}(2020{\natexlab{a}})\citenamefont {Agostini} \emph
  {et~al.}}]{Bx_nature_CNO}%
  \BibitemOpen
  \bibfield  {author} {\bibinfo {author} {\bibfnamefont {M.}~\bibnamefont
  {Agostini}} \emph {et~al.} (\bibinfo {collaboration} {Borexino
  Collaboration}),\ }\href {\doibase 10.1038/s41586-020-2934-0} {\bibfield
  {journal} {\bibinfo  {journal} {Nature}\ }\textbf {\bibinfo {volume} {587}},\
  \bibinfo {pages} {577–582} (\bibinfo {year}
  {2020}{\natexlab{a}})}\BibitemShut {NoStop}%
\bibitem [{\citenamefont {Agostini}\ \emph
  {et~al.}(2018{\natexlab{a}})\citenamefont {Agostini} \emph
  {et~al.}}]{Bx_Nature_2018}%
  \BibitemOpen
  \bibfield  {author} {\bibinfo {author} {\bibfnamefont {M.}~\bibnamefont
  {Agostini}} \emph {et~al.} (\bibinfo {collaboration} {Borexino
  Collaboration}),\ }\href {\doibase 10.1038/s41586-018-0624-y} {\bibfield
  {journal} {\bibinfo  {journal} {Nature}\ }\textbf {\bibinfo {volume} {562}},\
  \bibinfo {pages} {505} (\bibinfo {year} {2018}{\natexlab{a}})}\BibitemShut
  {NoStop}%
\bibitem [{\citenamefont {Abe}\ \emph {et~al.}(2016{\natexlab{a}})\citenamefont
  {Abe} \emph {et~al.}}]{superK_solar_neutrino_IV}%
  \BibitemOpen
  \bibfield  {author} {\bibinfo {author} {\bibfnamefont {K.}~\bibnamefont
  {Abe}} \emph {et~al.} (\bibinfo {collaboration} {Super-Kamiokande
  Collaboration}),\ }\href {\doibase 10.1103/PhysRevD.94.052010} {\bibfield
  {journal} {\bibinfo  {journal} {Physical Review D}\ }\textbf {\bibinfo
  {volume} {94}},\ \bibinfo {pages} {052010} (\bibinfo {year}
  {2016}{\natexlab{a}})}\BibitemShut {NoStop}%
\bibitem [{\citenamefont {Anderson}\ \emph {et~al.}(2019)\citenamefont
  {Anderson} \emph {et~al.}}]{SNOPlus_B8}%
  \BibitemOpen
  \bibfield  {author} {\bibinfo {author} {\bibfnamefont {M.}~\bibnamefont
  {Anderson}} \emph {et~al.} (\bibinfo {collaboration} {SNO+ Collaboration}),\
  }\href {\doibase 10.1103/PhysRevD.99.012012} {\bibfield  {journal} {\bibinfo
  {journal} {Physical Review D}\ }\textbf {\bibinfo {volume} {99}},\ \bibinfo
  {pages} {012012} (\bibinfo {year} {2019})}\BibitemShut {NoStop}%
\bibitem [{\citenamefont {Cleveland}\ \emph {et~al.}(1998)\citenamefont
  {Cleveland} \emph {et~al.}}]{Cleveland:1998nv}%
  \BibitemOpen
  \bibfield  {author} {\bibinfo {author} {\bibfnamefont {B.~T.}\ \bibnamefont
  {Cleveland}} \emph {et~al.} (\bibinfo {collaboration} {Homestake
  Collaboration}),\ }\href {\doibase 10.1086/305343} {\bibfield  {journal}
  {\bibinfo  {journal} {Astrophysical Journal}\ }\textbf {\bibinfo {volume}
  {496}},\ \bibinfo {pages} {505} (\bibinfo {year} {1998})}\BibitemShut
  {NoStop}%
\bibitem [{\citenamefont {Abdurashitov}\ \emph {et~al.}(2009)\citenamefont
  {Abdurashitov} \emph {et~al.}}]{SAGE:2009eeu}%
  \BibitemOpen
  \bibfield  {author} {\bibinfo {author} {\bibfnamefont {J.~N.}\ \bibnamefont
  {Abdurashitov}} \emph {et~al.} (\bibinfo {collaboration} {SAGE
  Collaboration}),\ }\href {\doibase 10.1103/PhysRevC.80.015807} {\bibfield
  {journal} {\bibinfo  {journal} {Physical Review C}\ }\textbf {\bibinfo
  {volume} {80}},\ \bibinfo {pages} {015807} (\bibinfo {year}
  {2009})}\BibitemShut {NoStop}%
\bibitem [{\citenamefont {Kaether}\ \emph {et~al.}(2010)\citenamefont
  {Kaether}, \citenamefont {Hampel}, \citenamefont {Heusser}, \citenamefont
  {Kiko},\ and\ \citenamefont {Kirsten}}]{Kaether:2010ag}%
  \BibitemOpen
  \bibfield  {author} {\bibinfo {author} {\bibfnamefont {F.}~\bibnamefont
  {Kaether}}, \bibinfo {author} {\bibfnamefont {W.}~\bibnamefont {Hampel}},
  \bibinfo {author} {\bibfnamefont {G.}~\bibnamefont {Heusser}}, \bibinfo
  {author} {\bibfnamefont {J.}~\bibnamefont {Kiko}}, \ and\ \bibinfo {author}
  {\bibfnamefont {T.}~\bibnamefont {Kirsten}},\ }\href {\doibase
  10.1016/j.physletb.2010.01.030} {\bibfield  {journal} {\bibinfo  {journal}
  {Physics Letters B}\ }\textbf {\bibinfo {volume} {685}},\ \bibinfo {pages}
  {47} (\bibinfo {year} {2010})}\BibitemShut {NoStop}%
\bibitem [{\citenamefont {Araki}\ \emph {et~al.}(2005)\citenamefont {Araki}
  \emph {et~al.}}]{KamLAND:2004mhv}%
  \BibitemOpen
  \bibfield  {author} {\bibinfo {author} {\bibfnamefont {T.}~\bibnamefont
  {Araki}} \emph {et~al.} (\bibinfo {collaboration} {KamLAND Collaboration}),\
  }\href {\doibase 10.1103/PhysRevLett.94.081801} {\bibfield  {journal}
  {\bibinfo  {journal} {Physical Review Letters}\ }\textbf {\bibinfo {volume}
  {94}},\ \bibinfo {pages} {081801} (\bibinfo {year} {2005})}\BibitemShut
  {NoStop}%
\bibitem [{\citenamefont {Ahmad}\ \emph {et~al.}(2002)\citenamefont {Ahmad}
  \emph {et~al.}}]{PhysRevLett.89.011301}%
  \BibitemOpen
  \bibfield  {author} {\bibinfo {author} {\bibfnamefont {Q.~R.}\ \bibnamefont
  {Ahmad}} \emph {et~al.} (\bibinfo {collaboration} {SNO Collaboration}),\
  }\href {\doibase 10.1103/PhysRevLett.89.011301} {\bibfield  {journal}
  {\bibinfo  {journal} {Physical Review Letters}\ }\textbf {\bibinfo {volume}
  {89}},\ \bibinfo {pages} {011301} (\bibinfo {year} {2002})}\BibitemShut
  {NoStop}%
\bibitem [{\citenamefont {Bellini}\ \emph {et~al.}(2011)\citenamefont {Bellini}
  \emph {et~al.}}]{Bellini:2011rx}%
  \BibitemOpen
  \bibfield  {author} {\bibinfo {author} {\bibfnamefont {G.}~\bibnamefont
  {Bellini}} \emph {et~al.} (\bibinfo {collaboration} {Borexino
  Collaboration}),\ }\href {\doibase 10.1103/PhysRevLett.107.141302} {\bibfield
   {journal} {\bibinfo  {journal} {Physical Review Letters}\ }\textbf {\bibinfo
  {volume} {107}},\ \bibinfo {pages} {141302} (\bibinfo {year}
  {2011})}\BibitemShut {NoStop}%
\bibitem [{\citenamefont {Grevesse}\ and\ \citenamefont {Sauval}(1998)}]{HZ}%
  \BibitemOpen
  \bibfield  {author} {\bibinfo {author} {\bibfnamefont {N.}~\bibnamefont
  {Grevesse}}\ and\ \bibinfo {author} {\bibfnamefont {A.}~\bibnamefont
  {Sauval}},\ }\href@noop {} {\bibfield  {journal} {\bibinfo  {journal} {Space
  Science Reviews}\ }\textbf {\bibinfo {volume} {85}},\ \bibinfo {pages} {161}
  (\bibinfo {year} {1998})}\BibitemShut {NoStop}%
\bibitem [{\citenamefont {Magg}\ \emph {et~al.}(2022)\citenamefont {Magg} \emph
  {et~al.}}]{HZ1}%
  \BibitemOpen
  \bibfield  {author} {\bibinfo {author} {\bibfnamefont {E.}~\bibnamefont
  {Magg}} \emph {et~al.},\ }\href {\doibase 10.1051/0004-6361/202142971}
  {\bibfield  {journal} {\bibinfo  {journal} {Astron. Astrophys.}\ }\textbf
  {\bibinfo {volume} {661}},\ \bibinfo {pages} {A140} (\bibinfo {year}
  {2022})}\BibitemShut {NoStop}%
\bibitem [{\citenamefont {Caffau}\ \emph {et~al.}(2010)\citenamefont {Caffau},
  \citenamefont {Ludwig}, \citenamefont {Steffen}, \citenamefont {Freytag},\
  and\ \citenamefont {Bonifacio}}]{LZ}%
  \BibitemOpen
  \bibfield  {author} {\bibinfo {author} {\bibfnamefont {E.}~\bibnamefont
  {Caffau}}, \bibinfo {author} {\bibfnamefont {H.-G.}\ \bibnamefont {Ludwig}},
  \bibinfo {author} {\bibfnamefont {M.}~\bibnamefont {Steffen}}, \bibinfo
  {author} {\bibfnamefont {B.}~\bibnamefont {Freytag}}, \ and\ \bibinfo
  {author} {\bibfnamefont {P.}~\bibnamefont {Bonifacio}},\ }\href {\doibase
  10.1007/s11207-010-9541-4} {\bibfield  {journal} {\bibinfo  {journal} {Solar
  Physics}\ }\textbf {\bibinfo {volume} {268}},\ \bibinfo {pages} {255}
  (\bibinfo {year} {2010})}\BibitemShut {NoStop}%
\bibitem [{\citenamefont {Asplund}\ \emph {et~al.}(2009)\citenamefont {Asplund}
  \emph {et~al.}}]{LZ1}%
  \BibitemOpen
  \bibfield  {author} {\bibinfo {author} {\bibfnamefont {M.}~\bibnamefont
  {Asplund}} \emph {et~al.},\ }\href@noop {} {\bibfield  {journal} {\bibinfo
  {journal} {Annu. Rev. Astron. Astrophys.}\ }\textbf {\bibinfo {volume}
  {47}},\ \bibinfo {pages} {481} (\bibinfo {year} {2009})}\BibitemShut
  {NoStop}%
\bibitem [{\citenamefont {Asplund}\ \emph {et~al.}(2021)\citenamefont {Asplund}
  \emph {et~al.}}]{LZ2}%
  \BibitemOpen
  \bibfield  {author} {\bibinfo {author} {\bibfnamefont {M.}~\bibnamefont
  {Asplund}} \emph {et~al.},\ }\href@noop {} {\bibfield  {journal} {\bibinfo
  {journal} {Astron. Astrophys.}\ }\textbf {\bibinfo {volume} {653}},\ \bibinfo
  {pages} {A141} (\bibinfo {year} {2021})}\BibitemShut {NoStop}%
\bibitem [{\citenamefont {Appel}\ \emph {et~al.}(2022)\citenamefont {Appel}
  \emph {et~al.}}]{Bx_improved_CNO}%
  \BibitemOpen
  \bibfield  {author} {\bibinfo {author} {\bibfnamefont {S.}~\bibnamefont
  {Appel}} \emph {et~al.} (\bibinfo {collaboration} {Borexino Collaboration}),\
  }\href {\doibase 10.1103/PhysRevLett.129.252701} {\bibfield  {journal}
  {\bibinfo  {journal} {Physical Review Letters}\ }\textbf {\bibinfo {volume}
  {129}},\ \bibinfo {pages} {252701} (\bibinfo {year} {2022})}\BibitemShut
  {NoStop}%
\bibitem [{\citenamefont {Agostini}\ \emph
  {et~al.}(2022{\natexlab{a}})\citenamefont {Agostini} \emph
  {et~al.}}]{Bx_CID_long}%
  \BibitemOpen
  \bibfield  {author} {\bibinfo {author} {\bibfnamefont {M.}~\bibnamefont
  {Agostini}} \emph {et~al.} (\bibinfo {collaboration} {Borexino
  Collaboration}),\ }\href {\doibase 10.1103/PhysRevD.105.052002} {\bibfield
  {journal} {\bibinfo  {journal} {Physical Review D}\ }\textbf {\bibinfo
  {volume} {105}},\ \bibinfo {pages} {052002} (\bibinfo {year}
  {2022}{\natexlab{a}})}\BibitemShut {NoStop}%
\bibitem [{\citenamefont {Agostini}\ \emph
  {et~al.}(2022{\natexlab{b}})\citenamefont {Agostini} \emph
  {et~al.}}]{Bx_CID_short}%
  \BibitemOpen
  \bibfield  {author} {\bibinfo {author} {\bibfnamefont {M.}~\bibnamefont
  {Agostini}} \emph {et~al.} (\bibinfo {collaboration} {Borexino
  Collaboration}),\ }\href {\doibase 10.1103/PhysRevLett.128.091803} {\bibfield
   {journal} {\bibinfo  {journal} {Physical Review Letters}\ }\textbf {\bibinfo
  {volume} {128}},\ \bibinfo {pages} {091803} (\bibinfo {year}
  {2022}{\natexlab{b}})}\BibitemShut {NoStop}%
\bibitem [{\citenamefont {Alimonti}\ \emph {et~al.}(2009)\citenamefont
  {Alimonti} \emph {et~al.}}]{bx_det}%
  \BibitemOpen
  \bibfield  {author} {\bibinfo {author} {\bibfnamefont {G.}~\bibnamefont
  {Alimonti}} \emph {et~al.} (\bibinfo {collaboration} {Borexino
  Collaboration}),\ }\href {\doibase
  https://doi.org/10.1016/j.nima.2008.11.076} {\bibfield  {journal} {\bibinfo
  {journal} {Nuclear Instruments and Methods in Physics Research Section A:
  Accelerators, Spectrometers, Detectors and Associated Equipment}\ }\textbf
  {\bibinfo {volume} {600}},\ \bibinfo {pages} {568} (\bibinfo {year}
  {2009})}\BibitemShut {NoStop}%
\bibitem [{\citenamefont {Bellini}\ \emph {et~al.}(2014)\citenamefont {Bellini}
  \emph {et~al.}}]{Bx_long_paper}%
  \BibitemOpen
  \bibfield  {author} {\bibinfo {author} {\bibfnamefont {G.}~\bibnamefont
  {Bellini}} \emph {et~al.} (\bibinfo {collaboration} {Borexino
  Collaboration}),\ }\href {\doibase 10.1103/PhysRevD.89.112007} {\bibfield
  {journal} {\bibinfo  {journal} {Physical Review D}\ }\textbf {\bibinfo
  {volume} {89}},\ \bibinfo {pages} {112007} (\bibinfo {year}
  {2014})}\BibitemShut {NoStop}%
\bibitem [{\citenamefont {Agostini}\ \emph
  {et~al.}(2018{\natexlab{b}})\citenamefont {Agostini} \emph
  {et~al.}}]{Bx_monte_carlo}%
  \BibitemOpen
  \bibfield  {author} {\bibinfo {author} {\bibfnamefont {M.}~\bibnamefont
  {Agostini}} \emph {et~al.} (\bibinfo {collaboration} {Borexino
  Collaboration}),\ }\href {\doibase 10.1016/j.astropartphys.2017.10.003}
  {\bibfield  {journal} {\bibinfo  {journal} {Astroparticle Physics}\ }\textbf
  {\bibinfo {volume} {97}},\ \bibinfo {pages} {136} (\bibinfo {year}
  {2018}{\natexlab{b}})}\BibitemShut {NoStop}%
\bibitem [{\citenamefont {Back}\ \emph {et~al.}(2012)\citenamefont {Back} \emph
  {et~al.}}]{Bx_calibration}%
  \BibitemOpen
  \bibfield  {author} {\bibinfo {author} {\bibfnamefont {H.}~\bibnamefont
  {Back}} \emph {et~al.} (\bibinfo {collaboration} {Borexino Collaboration}),\
  }\href {\doibase 10.1088/1748-0221/7/10/p10018} {\bibfield  {journal}
  {\bibinfo  {journal} {Journal of Instrumentation}\ }\textbf {\bibinfo
  {volume} {7}},\ \bibinfo {pages} {P10018} (\bibinfo {year}
  {2012})}\BibitemShut {NoStop}%
\bibitem [{\citenamefont {Abe}\ \emph {et~al.}(2016{\natexlab{b}})\citenamefont
  {Abe} \emph {et~al.}}]{super-kamiokande_IV}%
  \BibitemOpen
  \bibfield  {author} {\bibinfo {author} {\bibfnamefont {K.}~\bibnamefont
  {Abe}} \emph {et~al.} (\bibinfo {collaboration} {Super-Kamiokande
  Collaboration}),\ }\href {\doibase 10.1103/PhysRevD.94.052010} {\bibfield
  {journal} {\bibinfo  {journal} {Physical Review D}\ }\textbf {\bibinfo
  {volume} {94}},\ \bibinfo {pages} {052010} (\bibinfo {year}
  {2016}{\natexlab{b}})}\BibitemShut {NoStop}%
\bibitem [{\citenamefont {D'Agostini}(2003)}]{dagostini_bayes}%
  \BibitemOpen
  \bibfield  {author} {\bibinfo {author} {\bibfnamefont {G.}~\bibnamefont
  {D'Agostini}},\ }\href {\doibase 10.1088/0034-4885/66/9/201} {\bibfield
  {journal} {\bibinfo  {journal} {Reports on Progress in Physics}\ }\textbf
  {\bibinfo {volume} {66}},\ \bibinfo {pages} {1383} (\bibinfo {year}
  {2003})}\BibitemShut {NoStop}%
\bibitem [{\citenamefont {Agostini}\ \emph
  {et~al.}(2020{\natexlab{b}})\citenamefont {Agostini} \emph
  {et~al.}}]{Bx_CNO_sensitivity}%
  \BibitemOpen
  \bibfield  {author} {\bibinfo {author} {\bibfnamefont {M.}~\bibnamefont
  {Agostini}} \emph {et~al.} (\bibinfo {collaboration} {Borexino
  Collaboration}),\ }\href {\doibase 10.1140/epjc/s10052-020-08534-2}
  {\bibfield  {journal} {\bibinfo  {journal} {European Physics Journal C}\
  }\textbf {\bibinfo {volume} {80}},\ \bibinfo {pages} {1091} (\bibinfo {year}
  {2020}{\natexlab{b}})}\BibitemShut {NoStop}%
\bibitem [{\citenamefont {Capozzi}\ \emph {et~al.}(2021)\citenamefont
  {Capozzi}, \citenamefont {Di~Valentino}, \citenamefont {Lisi}, \citenamefont
  {Marrone}, \citenamefont {Melchiorri},\ and\ \citenamefont
  {Palazzo}}]{Capozzi:2021fjo}%
  \BibitemOpen
  \bibfield  {author} {\bibinfo {author} {\bibfnamefont {F.}~\bibnamefont
  {Capozzi}}, \bibinfo {author} {\bibfnamefont {E.}~\bibnamefont
  {Di~Valentino}}, \bibinfo {author} {\bibfnamefont {E.}~\bibnamefont {Lisi}},
  \bibinfo {author} {\bibfnamefont {A.}~\bibnamefont {Marrone}}, \bibinfo
  {author} {\bibfnamefont {A.}~\bibnamefont {Melchiorri}}, \ and\ \bibinfo
  {author} {\bibfnamefont {A.}~\bibnamefont {Palazzo}},\ }\href {\doibase
  10.1103/PhysRevD.104.083031} {\bibfield  {journal} {\bibinfo  {journal}
  {Physical Review D}\ }\textbf {\bibinfo {volume} {104}},\ \bibinfo {pages}
  {083031} (\bibinfo {year} {2021})}\BibitemShut {NoStop}%
\bibitem [{\citenamefont {Vescovi}\ \emph {et~al.}(2020)\citenamefont
  {Vescovi}, \citenamefont {Mascaretti}, \citenamefont {Vissani}, \citenamefont
  {Piersanti},\ and\ \citenamefont {Straniero}}]{Vescovi:2020wyz}%
  \BibitemOpen
  \bibfield  {author} {\bibinfo {author} {\bibfnamefont {D.}~\bibnamefont
  {Vescovi}}, \bibinfo {author} {\bibfnamefont {C.}~\bibnamefont {Mascaretti}},
  \bibinfo {author} {\bibfnamefont {F.}~\bibnamefont {Vissani}}, \bibinfo
  {author} {\bibfnamefont {L.}~\bibnamefont {Piersanti}}, \ and\ \bibinfo
  {author} {\bibfnamefont {O.}~\bibnamefont {Straniero}},\ }\href {\doibase
  10.1088/1361-6471/abb784} {\bibfield  {journal} {\bibinfo  {journal} {J.
  Phys. G}\ }\textbf {\bibinfo {volume} {48}},\ \bibinfo {pages} {015201}
  (\bibinfo {year} {2020})}\BibitemShut {NoStop}%
\bibitem [{\citenamefont {Bergstrom}\ \emph {et~al.}(2016)\citenamefont
  {Bergstrom}, \citenamefont {Gonzalez-Garcia}, \citenamefont {Maltoni},
  \citenamefont {Pena-Garay}, \citenamefont {Serenelli},\ and\ \citenamefont
  {Song}}]{Bergstrom:2016cbh}%
  \BibitemOpen
  \bibfield  {author} {\bibinfo {author} {\bibfnamefont {J.}~\bibnamefont
  {Bergstrom}}, \bibinfo {author} {\bibfnamefont {M.~C.}\ \bibnamefont
  {Gonzalez-Garcia}}, \bibinfo {author} {\bibfnamefont {M.}~\bibnamefont
  {Maltoni}}, \bibinfo {author} {\bibfnamefont {C.}~\bibnamefont {Pena-Garay}},
  \bibinfo {author} {\bibfnamefont {A.~M.}\ \bibnamefont {Serenelli}}, \ and\
  \bibinfo {author} {\bibfnamefont {N.}~\bibnamefont {Song}},\ }\href {\doibase
  10.1007/JHEP03(2016)132} {\bibfield  {journal} {\bibinfo  {journal} {Journal
  of High Energy Physics}\ }\textbf {\bibinfo {volume} {03}},\ \bibinfo {pages}
  {132} (\bibinfo {year} {2016})}\BibitemShut {NoStop}%
\end{thebibliography}%

\end{document}